\documentclass[fleqn,usenatbib]{mnras}

\usepackage{newtxtext,newtxmath}

\usepackage[T1]{fontenc}
\usepackage{ae,aecompl}
\usepackage{times}
\usepackage{enumitem}
\usepackage{times}
\usepackage{textcomp}
\usepackage{verbatim}
\usepackage{esdiff}

\newcommand{\msun}{{\,\rm M_\odot}}

\newcommand{\kms}{\,{\rm km}\,{\rm s}^{-1}}
\newcommand{\cm}{\,{\rm cm}}

\newcommand{\kpc}{\,{\rm kpc}}

\newcommand{\Mpc}{\,{\rm Mpc}}
\newcommand{\Gpc}{\,{\rm Gpc}}

\newcommand{\cpm}{\,{\rm cm}^2\,{\rm g}^{-1}}

\newcommand{\kev}{\,{\rm keV}}

\def\jcap{J. Cosmol.  Astropart. Phys.}
\def\aap{A\&A}
\def\apj{ApJ}

\def\apjl{ApJ}
\def\mnras{MNRAS}
\def\araa{ARA\&A}
\def\aj{AJ}

\def\physrep{Phys. Rep.}
\def\nat{Nature}

\def\apjs{ApJS}
\def\prd{Phys. Rev. D}

\DeclareRobustCommand{\VAN}[3]{#2}
\let\VANthebibliography\thebibliography
\def\thebibliography{\DeclareRobustCommand{\VAN}[3]{##3}\VANthebibliography}

\usepackage{graphicx}	
\usepackage{amsmath}	

\makeatletter
\newcommand{\rmnum}[1]{\romannumeral #1}
\newcommand{\Rmnum}[1]{\expandafter\@slowromancap\romannumeral #1@}

\renewcommand\paragraph{\@startsection{paragraph}{4}{\z@}{3.25ex\@plus1ex\@minus.2ex}{-1em}{\normalfont\it\normalsize}}

\title[X-ray morphology of clusters]{X-ray morphology of cluster-mass haloes in self-interacting dark matter}

\author[Shen et al.]{\parbox{17.5cm}{
Xuejian Shen,$^{1}$\thanks{E-mail: xshen@caltech.edu}
Thejs Brinckmann$^{2,3,4}$,
David Rapetti$^{5,6,7}$,
Mark Vogelsberger$^{8}$,
Adam Mantz$^{9}$,
Jes\'{u}s Zavala$^{10}$,
Steven W. Allen$^{9,11,12}$
\\
}\vspace{0.3cm}\\
$^{1}$ TAPIR, California Institute of Technology, Pasadena, CA 91125, USA \\
$^{2}$ Dipartimento di Fisica e Scienze della Terra, Universit\'a degli Studi di Ferrara, via Giuseppe Saragat 1, 44122 Ferrara, Italy\\
$^{3}$ Istituto Nazionale di Fisica Nucleare (INFN), Sezione di Ferrara, Via Giuseppe Saragat 1, 44122 Ferrara, Italy\\
$^{4}$ C.N. Yang Institute for Theoretical Physics and Department of Physics \& Astronomy, Stony Brook University, Stony Brook, NY 11794, USA\\
$^{5}$ NASA Ames Research Center, Moffett Field, CA 94035, USA \\
$^{6}$ Research Institute for Advanced Computer Science, Universities Space Research Association, Columbia, MD 21046, USA \\
$^{7}$ Center for Astrophysics and Space Astronomy, Department of Astrophysical and Planetary Science, University of Colorado, Boulder, CO 80309, USA \\
$^{8}$ Department of Physics, Kavli Institute for Astrophysics and Space Research, Massachusetts Institute of Technology, Cambridge, MA 02139, USA \\
$^{9}$ Kavli Institute for Particle Astrophysics and Cosmology, Stanford University, 452 Lomita Mall, Stanford, CA 94305, USA \\
$^{10}$ Centre for Astrophysics and Cosmology, Science Institute, University of Iceland, Dunhagi 5, 107 Reykjavik, Iceland \\
$^{11}$ Department of Physics, Stanford University, 382 Via Pueblo Mall, Stanford, CA 94305, USA\\
$^{12}$ SLAC National Accelerator Laboratory, 2575 Sand Hill Road, Menlo Park, CA 94025, USA
}

\date{Accepted XXX. Received YYY; in original form ZZZ}

\pubyear{2022}

\begin{document}

\label{firstpage}
\pagerange{\pageref{firstpage}--\pageref{lastpage}}
\maketitle

\begin{abstract}
We perform cosmological zoom-in simulations of $19$ relaxed cluster-mass haloes with the inclusion of adiabatic gas in the cold dark matter (CDM) and self-interacting dark matter (SIDM) models. These clusters are selected as dynamically relaxed clusters from a parent simulation with $M_{\rm 200} \simeq 1\operatorname{-}3\times 10^{15}\msun$. Both the dark matter and the intracluster gas distributions in SIDM appear more spherical than their CDM counterparts. Mock X-ray images are generated based on the simulations and are compared to the real X-ray images of $84$ relaxed clusters selected from the {\it Chandra} and {\it ROSAT} archives. We perform ellipse fitting for the isophotes of mock and real X-ray images and obtain the ellipticities at cluster-centric radii of $r\simeq 0.1\operatorname{-}0.2\,R_{\rm 200}$. The X-ray isophotes in SIDM models with increasing cross-sections are rounder than their CDM counterparts, which manifests as a systematic shift in the distribution function of ellipticities. Unexpectedly, the X-ray morphology of the observed non-cool-core clusters agrees better with SIDM models with cross-section $(\sigma/m)= 0.5\operatorname{-}1~\cpm$ than CDM and SIDM with $(\sigma/m)=0.1\cpm$. Our statistical analysis indicates that the latter two models are disfavored at the $68\%$ confidence level (as conservative estimates). This conclusion is not altered by shifting the radial range of measurements or applying temperature selection criterion. However, the primary uncertainty originates from the lack of baryonic physics in the adiabatic model, such as cooling, star formation and feedback effects, which still have the potential to reconcile CDM simulations with observations.  
\end{abstract}

\begin{keywords}
methods : numerical -- galaxies : clusters : general -- galaxies : clusters : intracluster medium -- cosmology : dark matter -- cosmology : theory
\end{keywords}



\section{Introduction}

The current paradigm for cosmological structure formation -- the cosmological constant plus cold dark matter ($\Lambda {\rm CDM}$) model -- has been remarkably successful in describing the large scale structure of the Universe~\citep[e.g.,][]{Blumenthal1984,Davis1985,Springel2005}. This model effectively assumes that the only dark matter interaction relevant for structure formation is gravity. It provides a clear picture of how small initial perturbations in the dark matter density field grow via gravitational instabilities and eventually collapse into dark matter haloes that harbour galaxy formation~\citep[e.g.,][]{White1991,Kauffmann1993,Cole2000,Springel2005,Hopkins2014,Vogelsberger2014_illustris,Schaye2015,Vogelsberger2020_review}. Despite the successes of the CDM model, new particle(s) that have the properties required to be CDM have not been discovered. Popular candidates for CDM (e.g., Weakly Interacting Massive Particles, WIMPs) have been the subject of fruitless extended searches, in particle colliders, and direct and indirect detection experiments, and a significant proportion of the WIMP parameter space has already been ruled out~\citep[e.g.,][]{Bertone2005,Bertone2010,Aprile2018,2018RPPh...81f6201R}. In addition, the $\Lambda$CDM paradigm has seen challenges on cosmological scales, most notably a tension between direct and indirect measurements of the Hubble parameter today, $H_0$~\citep[for a summary see, e.g.,][]{Verde:2019ivm}, but also a discrepancy between late time cosmic shear measurements and Cosmic Microwave Background measurements in the $\Omega_m-\sigma_8$ parameter plane~\citep[see, e.g.,][and references therein]{KiDS:2020suj,2021arXiv211006947L,2021arXiv210513544S,2021arXiv210513543A}, involving the total matter content of the universe and a measure for the amplitude of matter clustering. A plethora of attempts have been made to address the $H_0$ and $\Omega_m-\sigma_8$ discrepancies~\citep[for summaries see, e.g.,][]{DiValentino:2021izs,2021arXiv210710291S}, many involving more complex dark particle sectors with additional interactions between dark relics \citep[e.g.,][]{Cyr-Racine:2013jua,Archidiacono:2014nda,Archidiacono:2020yey,Baumann:2015rya,Forastieri:2015paa,Forastieri:2017oma,Forastieri:2019cuf,Lancaster:2017ksf,Choi:2018gho,Kreisch:2019yzn,Escudero:2019gvw,Blinov:2020hmc,Das:2020xke,RoyChoudhury:2020dmd,Brinckmann:2020bcn,Esteban:2021ozz,2021arXiv211100014A}, including models of dark matter interacting with light dark relics or dark radiation~\citep[e.g.,][]{vandenAarssen2012,2014PhRvD..90d3524B,Buen-Abad:2015ova,Buen-Abad:2017gxg,Cyr-Racine:2015ihg,Lesgourgues:2015wza,Archidiacono:2017slj,Archidiacono:2019wdp,DiValentino:2017oaw,Bose:2018juc,Ghosh:2019tab,Bohr:2020yoe,Becker:2020hzj,Munoz:2020mue,2021arXiv210806928G,2021arXiv211004024H,Mosbech:2020ahp}.

Meanwhile, astrophysical observations of dwarf galaxies have revealed outstanding (small-scale) challenges to the classical CDM picture~\citep[see a recent review by][]{Bullock2017}. For example, the {\it core-cusp} problem states that the central profiles of many dark matter dominated systems, such as dwarf spheroidal galaxies (dSphs) around the Milky Way and low-surface-brightness galaxies (LSBs), are likely to be cored as inferred by observations~\citep[e.g.,][]{Flores1994,Moore1994,deBlok2001,KDN2006,Gentile2004,Simon2005,Spano2008,KDN2011a,KDN2011b,Oh2011,Walker2011,Oh2015,Chan2015,Zhu2016}, in contrast to the universal cuspy central density profile found in dark-matter-only (DMO) simulations~\citep{Navarro1996,Navarro1997,Moore1999,Klypin2001,Navarro2004,Diemand2005}. On the other hand, the {\it too-big-to-fail} (TBTF) problem stems from the fact that a substantial population of massive and concentrated subhaloes identified in DMO simulations of Milky Way-mass hosts are incompatible with the stellar kinematics of the Local Group satellites~\citep{MBK2011,MBK2012,Kirby2014,Tollerud2014,Papastergis2015}. The TBTF problem for Milky Way satellites is currently a problem of diversity in the inner dark matter distribution, with some satellites being more compatible with cuspy haloes, and others with cored ones \citep{Read2019,Zavala2019}. This diversity (although not necessarily produced by the same effects) is also observed in more massive dwarf galaxies in the local environment \citep{Oman2015}.  

There are several physical mechanisms that are missing from DMO simulations, which are fundamental to alter the inner dark matter distribution in dwarf-size haloes, modifying the CDM predictions into ones that are more compatible with observations. Supernovae feedback is a well-known mechanism of cusp-core transformations via impulsive gravitational heating \citep{Pontzen2012} with an efficiency that depends on the energy, spatial distribution and timescales of the supernovae-driven episodic blowouts \citep[e.g.][]{Penarrubia2012,Burger2021}. For satellite galaxies, the environment of the host can play a major role in the matter distribution (both dark and baryonic) within the satellite depending on the evolution of its orbit. Tidal interaction with the Milky Way disk are predicted to have removed mass from the Milky Way satellites via tidal stripping and reduce the inner dark matter density via tidal shocking, particularly for orbits that pass close to the disk \citep[e.g.][]{Brooks2014,Fattahi2018,GK2019}. The combined effects of these {\it baryonic} mechanisms and the inclusion of observational effects/systematics can alleviate the CDM challenges mentioned above. They remain however, an active topic of debate, particularly the problem of diversity of the inner dark matter distribution in the population of dwarf galaxies, which remains poorly understood.

The CDM small-scale challenges combined with the unsuccessful search of CDM particles, has motivated several alternative dark matter models. Among them, self-interacting dark matter (SIDM) is an appealing category that has been proposed and studied for decades~\citep[e.g.,][]{Carlson1992, deLaix1995,Firmani2000,Spergel2000,Vogelsberger2013}. SIDM has the potential to solve many of the CDM small-scale astrophysical problems~\citep[see the review by][and references therein]{Tulin2018} and is well motivated by hidden dark sectors as extensions to the Standard Model of particle physics~\citep[e.g.,][]{Ackerman2009,Arkani-Hamed2009,Feng2009,Feng2010,Loeb2011,vandenAarssen2012,CyrRacine2013,Tulin2013,Cline2014}. 

Cosmological structure formation in SIDM has been investigated extensively since the first SIDM simulations in the early 2000's \citep{Yoshida2000,Dave2001,Colin2002}. The {\it thermal-averaged} local collision rate of dark matter is~\citep[e.g.,][]{Tulin2018,Shen2021}
\begin{align}
    \Gamma & = \rho_{\rm dm}\, \langle (\sigma/m)\, v_{\rm rel}\rangle \nonumber \\
    & \simeq 0.15\,{\rm Gyr}^{-1} \left(\dfrac{\rho_{\rm dm}}{0.1\msun/{\rm pc}^3}\right)\,\left(\dfrac{\sigma/m}{0.1\cpm} \right)\,\left(\dfrac{\sigma_{\rm v}}{50\kms} \right),
    \label{eq:rate_colli}
\end{align}
where $(\sigma/m)$ is the self-interaction cross-section (per unit mass, at the velocity scale of interest), $\rho_{\rm dm}$ is the dark matter mass density and $\sigma_{\rm v}$ is the three-dimensional velocity dispersion of dark matter particles. When $(\sigma/m)\ll 0.1\cpm$, the cosmological impact of self-interactions would be negligible (as can be seen by comparing $\Gamma$ with the Hubble expansion rate). High-resolution DMO simulations of SIDM have found that a self-interaction cross-section of $\sim 0.1\operatorname{-}1\cpm$ could solve the {\it core-cusp} and TBTF problems in dwarf galaxies~\citep[e.g.,][]{Vogelsberger2012, Rocha2013, Zavala2013, Elbert2015, Dooley2016}. SIDM with comparable cross-sections can also explain the diversity of the inner dark matter distribution as given by stellar kinematics in the Local Group satellites~\citep{Tollerud2014,SGK2019,Sameie2020} and rotation curves in gas-rich dwarfs in the local environment~\citep[e.g.,][]{Kamada2017,Creasey2017,Kaplinghat2019,Omid2020}. Velocity-dependent SIDM models with large cross-sections at the scale of dwarf galaxies are equally successful~\citep[e.g.,][]{Zavala2013,Zavala2019,Turner2021} and can actually produce a distinct subhalo population with a bimodal behaviour: cuspy dark matter suhaloes in the smallest dwarfs caused by the gravothermal collapse mechanism \citep[e.g.][]{Balberg2002,Koda2011} and cored dark matter subhaloes for the larger dwarfs. Moreover, exotic scenarios involving exothermic or endothermic (dissipative) processes from inelastic scattering have been considered to evaporate the Milky Way satellites \citep{Vogelsberger2019} or seed supermassive black holes at high redshift~\citep[e.g.,][]{Choquette2019,Xiao2021}.

Assuming $(\sigma/m)$ is velocity-independent, Equation~\ref{eq:rate_colli} implies that the signature of SIDM will be stronger in systems with higher densities and velocity dispersions, so naturally the most stringent constraints on SIDM come from massive galaxy clusters. For instance, constraints around $0.4\operatorname{-}2\cpm$ ($95\%$ confidence level) have been obtained from the lack of a spatial offset between the total mass peak and galaxy centroid~\citep[e.g.,][]{Clowe2006, Randall2008, Kahlhoefer2015, Harvey2015, Wittman2018} in merging bullet-like clusters, or the strength of wobbles of the bright central galaxy~\citep[BCG;][]{Harvey2019}. The robustness of these constraints is still under debate due to the difficulty in measuring and interpreting observables given the complexity of the baryonic physics and their interplay with the SIDM physics~\citep[e.g.,][]{Vogelsberger2014,Kaplinghat2016, Robles2017, Elbert2018, Fitts2019, Robertson2019}. 

On the other hand, dark matter halo shapes are a viable alternative avenue to constrain SIDM with several stuides made in the past. For example, \citet{Miralda2002} argued that dark matter haloes should be spherical inside the radius where dark matter particles would collide with each other once during a Hubble time on average. Based on the shape of the galaxy cluster MS 2137-23 as inferred from strongly gravitationally-lensed arcs, \citet{Miralda2002} obtained a stringent constraint on the SIDM cross-section, $\sigma/m \lesssim 0.02 \cpm$. Such a strong constraint was later shown to be incorrect by \citet{Peter2013}, by demonstrating that one collision event of dark matter particles on average is not enough to make haloes completely spherical and that projection effects need to be properly considered to interpret observations. As a result, the constraint on SIDM was weakened to $\sigma/m \lesssim 1\cpm$. In recent years, high resolution X-ray imaging data have provided rich information on the intracluster gas over a large dynamical range and have been used to infer the shapes of matter distributions in galaxy clusters~\citep[e.g.,][]{Hashimoto2007, Kawahara2010}, which has direct implications for SIDM constraints. In addition, X-ray morphological studies are also a powerful tool to assess the dynamical state of the intracluster medium (ICM). Samples of massive relaxed clusters have been identified through quantitative studies of the morphology of X-ray selected clusters~\citep[e.g.,][]{Jeltema2005,Santos2008,Bohringer2010,Nurgaliev2013,Rasia2013,Mantz2015}. These clusters are ideal to compare to simulated counterparts in near equilibrium states in order to place significant constraints on SIDM based on their shapes. 

In this paper, we perform a series of cosmological zoom-in simulations of cluster-mass haloes in SIDM, building upon the DMO work of \citet{Brinckmann2018} by including adiabatic gas. We then compare these simulated clusters to $84$ massive observed clusters selected by~\cite{Mantz2014,Mantz2015} and derive constraints for SIDM models through the analyses of cluster X-ray morphology. The paper is organized as follows: details of the simulations are introduced in Section~\ref{sec:sim}, while the observational samples are introduced in Section~\ref{sec:obs}. In Section~\ref{sec:method}, we discuss the modelling of the X-ray emission and the generation of mock images for the simulated clusters. Details of the morphological analysis of the mock and real X-ray images are also discussed in this section. The results of the paper are presented in Section~\ref{sec:results} and are discussed further in Section~\ref{sec:discussion}. Finally, we summarize and conclude in Section~\ref{sec:conclusion}.

\section{Simulations}
\label{sec:sim}

The analysis in this paper is based on a suite of cosmological zoom-in simulations of cluster-mass haloes (with the DMO version presented in \citealt{Brinckmann2018,Sokolenko2018}). The simulations are performed using the moving-mesh code {\sc Arepo}~\citep{Springel2010} with the inclusion of adiabatic gas. The code employs the tree-particle-mesh (Tree-PM) algorithm for gravity and a finite-volume/Godunov scheme for hydrodynamics on an unstructured, moving Voronoi mesh. The haloes for zoom-in simulations were selected as dynamically relaxed systems from a large $1 (\Gpc/h)^{3}$ parent simulation with an effective resolution of $512^{3}$ dark matter particles \citep[see][for details on the relaxation criteria used]{Brinckmann2018}. The zoom-in simulations have an effective resolution of $4096^3$ dark matter particles in the high resolution regions, which are surrounded by regions of intermediate resolution and finally low resolution regions with an effective resolution of $256^3$ particles. For the high resolution region, the effective Plummer equivalent gravitational softening length of dark matter is $\epsilon=5.4 \kpc /h$ and the dark matter particle mass resolution is $m_{\rm dm} = 1.07 \times 10^{9} \msun/h$.

Dark matter self-interactions were simulated in a Monte Carlo fashion using the module developed in \citet{Vogelsberger2012,Vogelsberger2016}, assuming isotropic and elastic scattering. In this work, we only study the case of a constant self-interaction cross-section, and in particular we perform simulations for three cases: $(\sigma/m) = 0.1 \cpm$ (SIDM-c0.1), $(\sigma/m) = 0.5 \cpm$ (SIDM-c0.5), $(\sigma/m) = 1 \cpm$ (SIDM-c1), in addition to the CDM case for comparison. Our simulations use the cosmological parameters originally adopted in \citet{Brinckmann2018}: $\Omega_{\rm m}=0.315$, $\Omega_{\Lambda}=0.685$, $\Omega_{\rm b}=0.049$, $h=0.673$, $\sigma_{\rm 8} =0.83$ and $n_{\rm s} =0.96$, which are consistent with Planck results~\citep{Planck2016}.

Compared to the DMO version of the simulations in \citet{Brinckmann2018,Sokolenko2018}, our simulations introduce adiabatic gas cells, which are generated in the initial conditions by splitting dark matter particles, with the mass ratio between gas and dark matter particles set initially by the universal baryon fraction. The gas cells (as Voronoi meshes) are regularized by their masses or face solid angles and are allowed to be split or merged. The baryonic mass resolution in the final halo is roughly the initial gas cell mass, $m_{\rm b} \simeq m_{\rm dm}\, \Omega_{\rm b}/(\Omega_{\rm m}-\Omega_{\rm b}) \simeq 0.18\, m_{\rm dm} \simeq 2 \times 10^{8} \msun/h$. The spatial resolution of hydrodynamics is roughly the cell equivalent size (the radius of the sphere with the average volume of the cells) $h_{\rm b} = 4.8 \kpc/h \times (\rho_{\rm b}/10^{5} \msun/\kpc^{3})^{-1/3}$, where $10^{5} \msun/\kpc^{3}$ is the typical gas density at cluster centers in our simulations. The gravitational softening length of adiabatic gas is chosen to be the same as that of dark matter, i.e., $\epsilon_{\rm gas}=5.4 \kpc /h$ and the adiabatic index of gas is chosen to be $5/3$.

The main target haloes are identified in the zoom-in regions and the dark matter particles or gas cells are assigned to the main target haloes using the Friends-of-Friends (FoF) algorithm. The virial mass and radius of each halo are defined based on the density criterion, $200$ times the critical density at $z=0$\footnote{Some cluster studies adopt instead the redshift-dependent overdensity criterion from \citet{Bryan1998} which gives $\Delta_{\rm c}(z=0)\simeq 100$. This could lead to about $30\%$ ($10\%$) increase in the virial radius (mass).}, and are therefore referred to as $M_{\rm 200}$ and $R_{\rm 200}$, respectively. The virial temperature is defined as $T_{\rm vir} = (\mu m_{\rm p}/2k_{\rm B}) GM_{\rm 200}/R_{\rm 200}$, where $m_{\rm p}$ is the proton mass, $k_{\rm B}$ is the Boltzmann constant and $\mu$ is the mean molecular weight that takes the value $0.59$ (see also Equation~\ref{eq:molecular_weight}).

The convergence radius of collisionless particles can be calculated using the \citet{Power2003} criterion. \citet{Power2003} argued that the artificial central ``flattening'' of dark matter profiles is driven by two-body relaxation, and that robust results should be obtained outside the radius where the relaxation time is comparable to the Hubble time. This is equivalent to the criterion
\begin{equation}
    \dfrac{\sqrt{200}}{8} \dfrac{N(r)}{\ln{(N(r))}} \left( \dfrac{\bar{\rho}(r)}{\rho_{\rm crit}} \right)^{-1/2} \geq 0.6,
    \label{eq:power}
\end{equation}
where $N(r)$ is the number of particles within a radius $r$, $\rho_{\rm crit}$ is the critical density of the Universe at $z=0$ and $\bar{\rho}(r)$ is the average density within $r$. We evaluate the convergence radius for each of our simulations based on this criterion and the obtained values are listed in Table~\ref{tab:sims}. On the other hand, the convergence of the hydrodynamical properties of the gas is more complicated and depends on the numerical method employed. In Section~\ref{sec:results-denpro}, we will explicitly check how the hydrodynamical properties of the gas in our simulations are resolved and discuss the issue of convergence. 

The typical virial mass of the simulated haloes is $M_{\rm 200} \simeq (1\,\operatorname{-}\,3)\times10^{15} \msun$ and the typical size is $R_{\rm 200} \simeq 2\,\operatorname{-}\,3 \Mpc$. The detailed properties of all the simulated haloes are listed in Table~\ref{tab:sims}. 

\begin{table}
    \centering
    \begin{tabular}{p{0.06\textwidth}|p{0.09\textwidth}|p{0.07\textwidth}|p{0.07\textwidth}|p{0.07\textwidth}}
        \hline
        Halo$^{\rm a}$ & $M^{\rm cdm}_{\rm 200}$ & $R^{\rm cdm}_{\rm 200}$ & $T^{\rm cdm}_{\rm vir}$ & $R_{\rm conv}$$^{\rm b}$ \\
        name & \,\,\,\,$[10^{15}\msun]$ & $[{\rm Mpc}]$ & $[10^{7}{\rm K}]$ & $[{\rm kpc}]$ \\
        \hline
        \hline
    \end{tabular}
   
    \begin{tabular}{p{0.06\textwidth}|p{0.09\textwidth}|p{0.07\textwidth}|p{0.07\textwidth}|p{0.07\textwidth}}
        halo11  & 2.91 & 3.03 & 14.7  & 35.9 \\
        halo39  & 1.47 & 2.41 & 9.29  & 33.7\\
        halo43  & 1.54 & 2.45 & 9.61  & 35.0\\
        halo55  & 1.38 & 2.36 & 8.94  & 35.0\\
        halo83  & 1.52 & 2.43 & 9.50  & 35.8\\
        halo84  & 1.66 & 2.51 & 10.1  & 34,7\\ 
        halo92  & 1.32 & 2.33 & 8.68  & 34.3\\
        halo102 & 1.39 & 2.37 & 8.98  & 33.8\\
        halo128 & 1.40 & 2.37 & 9.00  & 35,6\\
        halo136 & 1.10 & 2.18 & 7.65  & 33.0\\
        halo144 & 1.50 & 2.42 & 9.43  & 34.1\\
        halo159 & 1.25 & 2.28 & 8.35  & 35.5\\
        halo162 & 1.34 & 2.34 & 8.75  & 34.8\\
        halo165 & 1.26 & 2.28 & 8.38  & 36.1\\
        halo171 & 1.30 & 2.31 & 8.59  & 34.8\\
        halo194 & 1.54 & 2.45 & 9.61  & 35.3\\
        halo210 & 1.15 & 2.22 & 7.88  & 36.1\\
        halo215 & 1.24 & 2.27 & 8.30  & 36.0\\
        halo217$^{\rm c}$ & 1.29 & 2.57 & 7.68 & 38.9\\
        \hline
    \end{tabular}
    \caption{Simulated cluster-mass haloes in the suite.
    \newline \hspace{\textwidth}
    (\textbf{a}) Each halo is simulated in CDM, SIDM-c0.1, SIDM-c0.5 and SIDM-c1. The bulk properties of these haloes are indistinguishable in different dark matter models, so we only list the properties in the CDM simulations here.
    \newline
    (\textbf{b}) The radius of convergence of dark matter properties (based on the \citealt{Power2003} criterion discussed in Section~\ref{sec:sim}). We present the maximum convergence radius for simulations in all four dark matter models as a conservative estimate.
    \newline
    (\textbf{c}) Due to a technical issue, the simulation was stopped at $z\simeq 0.18$ instead of $z=0$. We approximate the $z=0$ results with this snapshot.}
    \label{tab:sims}
\end{table}

\section{Observational samples}
\label{sec:obs}

The observational samples we use consist of relaxed galaxy clusters as selected in \citet{Mantz2014,Mantz2015} using three morphological indicators, symmetry, peakiness and alignment, of cluster X-ray images. \citet{Mantz2015} developed a symmetry--peakiness--alignment (SPA) criterion for relaxation and applied this analysis to a large sample of galaxy clusters with archival {\it Chandra} and {\it ROSAT} observations, which resulted in $40$ relaxed clusters at $z\lesssim 1$. Each of these clusters has the cleaned science image, the blank-sky event file and an appropriate exposure map, along with the blank-sky normalization factor and its statistical error, which all serve as input to the morphological algorithm. Details of the sample selection, data reduction and post-processing can be found in \citet{Mantz2014,Mantz2015}. The typical ICM temperature of these clusters is $5\operatorname{-}10 \kev$ (about $5\operatorname{-}10\times 10^{7}\,{\rm K}$), which is in good agreement with the virial temperatures of the simulated clusters listed in Table~\ref{tab:sims}. The original peakiness criterion, however, preferentially selects clusters with cool cores, which indicate strong radiative cooling processes at cluster centers. Cooling and the subsequent star formation as well as supernovae and active galactic nuclei (AGN) feedback could significantly impact the structure of the central halo. Since our simulations do not capture these processes, we specifically select another set of clusters that meet the symmetry--alignment criterion but not the peakiness criterion, referred to as the ``non--peak'' clusters, while the original set of SPA selected clusters are instead referred to as the ``peaky'' clusters. The new set of ``non--peaky'' clusters consists of $44$ relaxed clusters. We will perform analyses on both sets of clusters to study the potential impact of cluster cool cores on X-ray morphology.

\begin{figure*}
    \centering
    \includegraphics[width=0.49\textwidth]{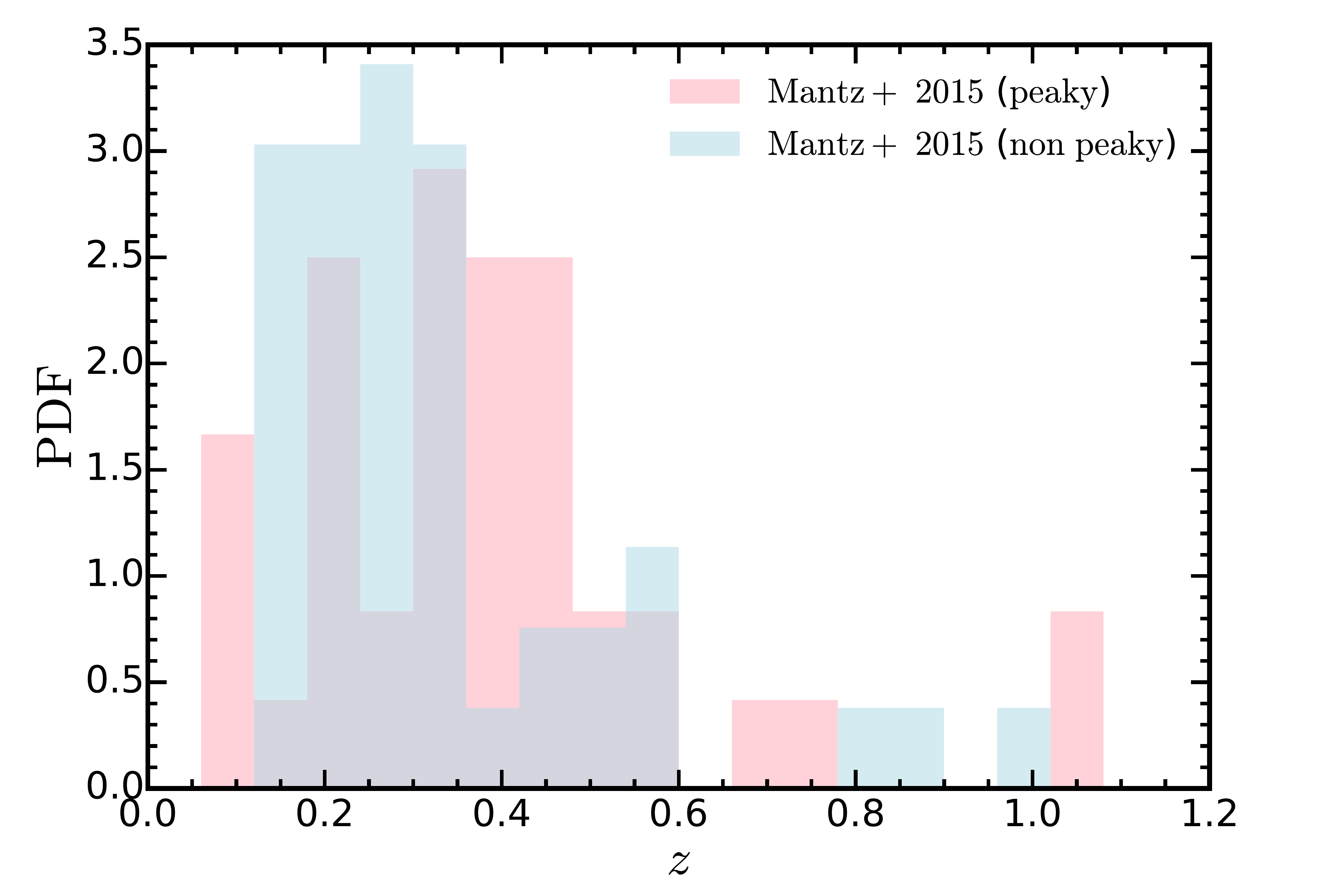}
    \includegraphics[width=0.49\textwidth]{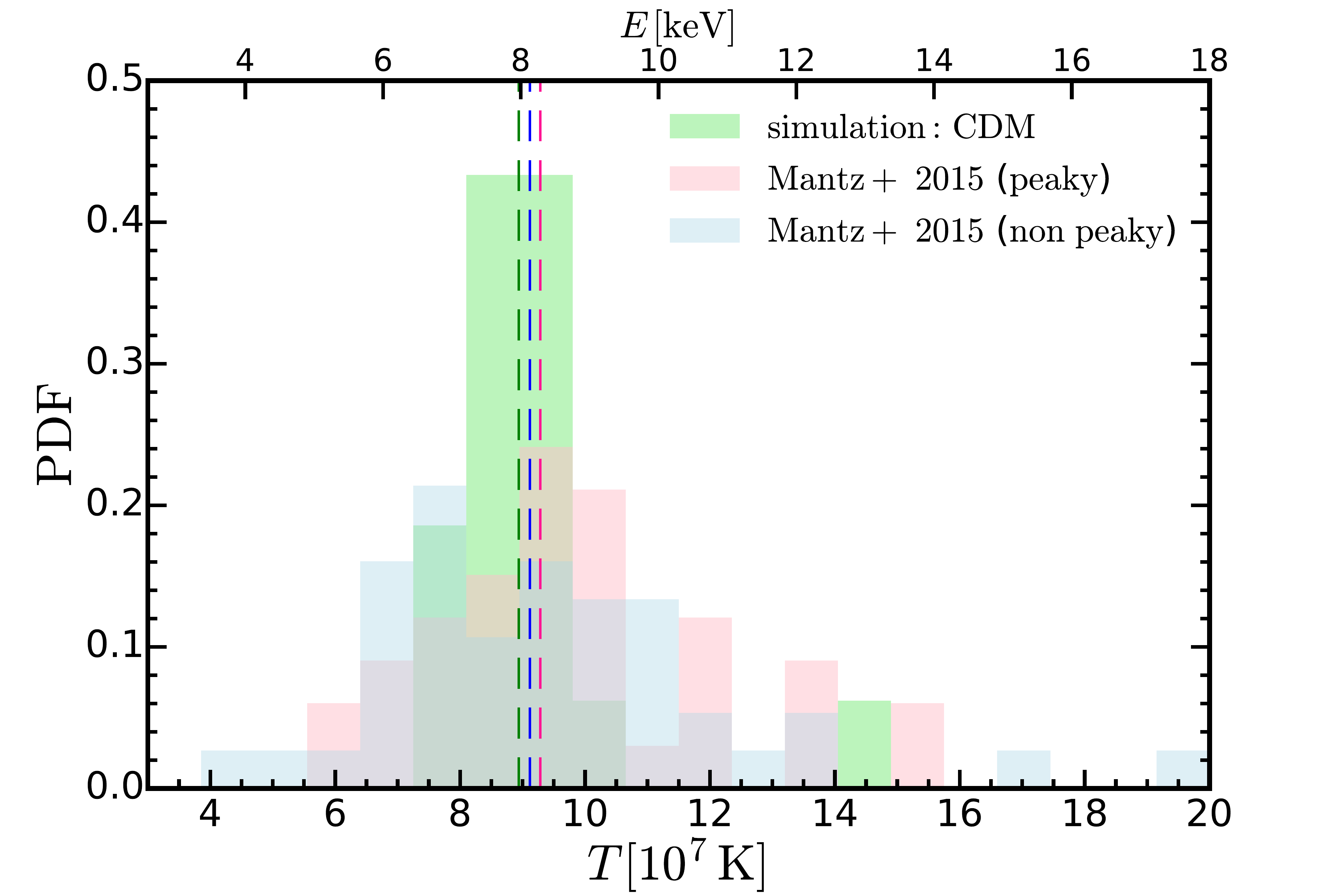}
    \caption{Redshift and temperature distributions of the observed clusters and temperature distribution for the simulated clusters. {\it Left:} Redshift distribution of the observed clusters. The distributions of the ``peaky'' and ``non-peaky'' samples are shown in red and blue. For both distributions, most of the clusters fall in the range $0.1\lesssim z \lesssim 0.5$, with a few outliers out to $z\sim 1$. {\it Right:} Temperature distribution of the observed (red and blue) and the simulated clusters (green). The median temperature of each sample is shown by a corresponding vertical dashed line. On average, the temperatures of simulated clusters are fairly consistent with the observed samples, despite having a smaller dispersion in temperature.}
    \label{fig:ztdis}
\end{figure*}

In Figure~\ref{fig:ztdis}, we show the redshift (left panel) and temperature distributions (right panel) of the observed clusters and the temperature distribution for the simulated clusters (right panel). For simulations, the ICM gas temperature is approximated as the virial temperature $T_{\rm vir}$, which is tested to be close to the X-ray surface-brightness weighted temperature of ICM gas. Most of the observed clusters, either the ``peaky'' or the ``non-peaky'' ones, fall in the redshift range $0.1\lesssim z\lesssim 0.5$. The median temperatures of the observational samples and the simulated clusters match reasonably well, but the observational samples show larger dispersion in temperature.

\section{Methods}
\label{sec:method}

\citet{Brinckmann2018} found that halo shapes are more sensitive to dark matter self-interactions at larger radii than spherically-averaged density profiles. Signatures of SIDM can be found in halo shapes out to the radii where density profiles already converge to the CDM prediction. The radial range of $10\operatorname{-}20\%\, R_{\rm 200}$ was found to be a suitable range where substantial differences between SIDM and CDM are observed in DMO simulations, and where it was speculated that the impact of complicated baryonic physics in the central galaxy would be limited. The primary goal of the present work is to have a more direct comparison of the halo morphology from simulations and observations, through more realistic modelling of the X-ray emission from simulated clusters and two-dimensional shape analysis of mock/real X-ray images in the radial range of interest.

\subsection{Mock X-ray images}
\label{sec:method-XrayMock}

We begin by generating the X-ray spectrum for every gas cell in each of the haloes based on a table of spectral templates. The templates are calculated using the Astrophysical Plasma Emission Code \citep[APEC][]{Smith2001} model implemented in the {\sc PyAtomDB} code\footnote{\href{https://atomdb.readthedocs.io/en/master/index.html}{https://atomdb.readthedocs.io/en/master/index.html}}, which utilized the atomic data from AtomDB v3.0.9 \citep[last described in][]{Foster2012}. The model gives the emission spectrum of collisional-ionized diffuse gas in equilibrium with a given temperature and metal abundance pattern. The temperature of a gas cell from the simulations is calculated as
\begin{equation}
    T = \dfrac{(\gamma -1 ) U \mu m_{\rm p}}{k_{\rm B}},
\end{equation} 
where $U$ is the internal energy of the gas cell, $\gamma=5/3$ is the assumed adiabatic index and $\mu$ is the mean molecular weight, which can be calculated as
\begin{equation}
    \mu = \dfrac{4}{1+3X_{\rm H}+4X_{\rm H}x_{\rm e}},
    \label{eq:molecular_weight}
\end{equation}
where $X_{\rm H}=0.76$ is the hydrogen mass fraction in the Universe and $x_{\rm e}\,(\equiv n_{\rm e}/n_{\rm H})$ is the electron abundance, assumed to be $1.17$~\citep{AG1989}. The abundance pattern is set to solar values following \citet{AG1989}, while the ICM metallicity is set to $0.25$ $Z_{\odot}$~\citep[e.g.,][]{McDonald2016,Mantz2017}. We note that, for hot intracluster gas as considered here ($ T \gtrsim 10^{7}\,{\rm K}$), the emission is dominated by thermal Bremsstrahlung and it is insensitive to details of the abundance pattern. Then, we account for galactic absorption with the photoelectric absorption cross-section given by \citet{Morrison1983}, assuming a fixed galactic hydrogen column density of $N_{\rm H} = 2\times 10^{20} \cm^{-2}$. This effectively decreases the rest-frame soft X-ray luminosity by only $\sim 3\%$. The energy range and resolution of the spectra depend on the desired instrument. For example, an instrument similar to Chandra ACIS-I has an energy range of $0.5\,\operatorname{-}\,10 \kev$ with an energy resolution of $150\,{\rm eV}$. For our templates, we adopt energy bins with high resolution $10\,{\rm eV}$ across $0.1\,\operatorname{-}\,100 \kev$. These spectral templates describe the energy emitted per unit time in each energy bin, $f(E,T)$, normalized by the emission measure. Assuming the size of the cluster is much smaller than the cosmological distances involved, the observed X-ray flux (per unit energy per unit area and per unit time) can be calculated as
\begin{align}
    f^{\rm obs}(E_{\rm obs}) & = \dfrac{(1+z)}{4\pi D_{\rm L}^2}\int_{\rm l.o.s.} f^{\rm rst}\big(E_{\rm rst}, T\big)\, n_{\rm e}\, n_{\rm H}\, {\rm d}V \nonumber \\
    & = \dfrac{(1+z)}{4\pi D_{\rm L}^2}\int_{\rm l.o.s.} f^{\rm rst}\big((1+z)E_{\rm obs}, T\big)\,n_{\rm e}\, n_{\rm H}\,{\rm d}V,
\end{align}
where $D_{\rm L}$ is the luminosity distance, $n_{\rm H}$ ($n_{\rm e}$) is the hydrogen (electron) number density, ``obs'' and ``rst'' refer to the observer's frame and the rest frame, respectively. The integration is performed along the line of sight. If we consider the integrated luminosity in an energy band in the observer's frame, we obtain
\begin{align}
    F^{\rm obs} & = \int_{E_{\rm min}}^{E_{\rm max}} f^{\rm obs}(E_{\rm obs}) {\rm d}E_{\rm obs} \nonumber \\
    & =  \dfrac{1}{4\pi D_{\rm L}^2} \int_{\rm l.o.s.}  n_{\rm e}\, n_{\rm H}\, {\rm d}V \, \int_{(1+z)E_{\rm min}}^{(1+z)E_{\rm max}} f^{\rm rst}\big(E_{\rm rst}, T\big)\,{\rm d}E_{\rm rst},
    \label{eq:Fobs}
\end{align}
where we choose $E_{\rm min},E_{\rm max}=0.6,2 \kev$ for the soft X-ray band images, and for simulated haloes (evolved to $z=0$) we assume a small dummy ``emission'' redshift of $0.03$, which does not have any real impact on the flux except for a constant normalization change. In practice, we choose to evaluate the integral over energy in Equation~\ref{eq:Fobs} first, solely based on the spectral templates. Then, we evaluate the line-of-sight integral based on the particle information obtained from simulations.

Finally, for each pixel with physical side length $L_{\rm p}$, the surface brightness can be calculated as
\begin{equation}
    S.B.({\rm pixel}) \simeq \dfrac{ F^{\rm obs} }{ (L_{\rm p}/D_{\rm A})^2 },
\end{equation}
where $D_{\rm A}$ is the angular diameter distance. For an annulus with a surface area $A$ and cluster-centric radius $r$, the surface brightness profile of a simulated cluster can be calculated in a similar way
\begin{equation}
    S.B.(r) \simeq \dfrac{ F^{\rm obs} D_{\rm A}^2 }{ A(r) }.
\end{equation}

For each simulated cluster, we pick $12$ viewing angles that correspond to the $12$ vertices of the $N=1$ Healpix sphere~\citep{Gorski2005} oriented in the simulation coordinates. Then, for each viewing angle, we generate an X-ray image of the cluster with a physical side length of $L=0.6\,R_{\rm 200}$ and $N_{\rm p}=1024$ pixels on each side, following the steps described above. Gas cells are binned in pixels and the X-ray surface brightness in the soft X-ray band ($0.6-2\kev$) is calculated for each pixel. We note that the equivalent size $h_{\rm b}$ of the gas cell could be larger than the physical size of the pixels. So the X-ray emitting gas cells should be considered as smoothed distributions of emitting material rather than discrete particles. As an approximate correction for this effect\footnote{In principle, the gas cells should be smoothed before being binned in pixels and used in flux calculations. For our application, this is equivalent to smoothing after the images are generated. The argument is supported by the following estimations: The typical displacement of particle coordinates to the pixel center scales as $1/n^{1/2}\, [{\rm pixel}]$, where $n$ is the number of particles projected in a pixel $\sim (L/h_{\rm b})^{3}/N^{2}_{\rm p}$. For reference, the smoothing kernel bandwidth is $N_{\rm p}\,h_{\rm b}/L\, [{\rm pixel}]$. The ratio of the two is a constant $\sim (h_{\rm b}/L)^{1/2} \sim 0.07$, which corresponds to $\sim 0.03$ in the logarithm of the flux, and is therefore small enough to be neglected.}, the images are convolved with a Gaussian kernel with bandwidth $h_{\rm b}$.

\subsection{Shape analysis of X-ray isophotes}
\label{sec:method-2dshape}

Based on the mock X-ray images created from the simulations, we use the {\sc Isophote} package in the {\sc Photutils} code~\citep{larry_bradley_2020} to perform ellipse fitting of isophotes using the iterative algorithm introduced in \citet{Jedrzejewski1987}. Each isophote is fitted for a pre-defined semi-major axis length. The algorithm starts from a first guess of the elliptical isophote, defined by approximate values of center coordinates, ellipticity ($e$) and position angle ($\phi$). The ellipticity is defined as
\begin{equation}
    e = 1 - \dfrac{b}{a},
    \label{eq:e_definition}
\end{equation}
where $a$ and $b$ are the semi-major and semi-minor axes of the ellipse, respectively. Then the fitting is done recursively to minimize the intensity variations of pixels along the elliptical path. For the first guess, we choose the semi-major axis to be $15\%\,R_{\rm 200}$ (the median of the radial range of interest) and set the center of the ellipse as the cluster center. We then derive the first guess of the ellipticity and position angle by recursively doing isophote fitting at the semi-major axis of $15\%\,R_{\rm 200}$, until the ellipticity and position angle are converged ($\Delta e<0.03$, $\Delta \phi < 0.03 \times 2\pi$). After fitting the ellipse that corresponds to a given value of the semi-major axis, the axis length is incremented (or decremented) following a pre-defined rule and the fitting procedure is repeated again at the new semi-major axis. The first guess for the ellipse parameters is taken from the previously fitted ellipse with the closest semi-major axis length to the current one. The fitting will be terminated when either the maximum acceptable relative error in the local radial intensity gradient is reached or a significant fraction of pixels on the ellipse lie outside the image. We define the effective radius of a fitted isophote as the geometric mean of the semi-major and semi-minor axes, $r_{\rm eff} = \sqrt{(a^2 + b^2)/2}$, and the results can be translated into ellipticity values as a function of $r_{\rm eff}$. To get a measure of the ellipticity in the radial range of interest, we compute the average ellipticity at $10\operatorname{-}20\% \,R_{\rm 200}$. 

For the X-ray images from observations, we use the {\sc Spa} code developed in \citet{Mantz2014,Mantz2015}, which was used for the original sample selection and morphological analysis, to perform isophotes identification and ellipse fitting. Along with the cleaned science image, the algorithm takes the exposure map of observations, the sky background noise, the blank-sky normalization factor and its statistical error as inputs. We refer readers to \citet{Mantz2015} for a detailed description of the algorithm. To standardize the surface brightness of clusters, the code motivated a redshift- and temperature-dependent scaling of the surface brightness based on the self-similar model of \citet{Kaiser1986} \citep[see also ][]{Santos2008}. The surface brightness is normalized in units of
\begin{align}
    \label{eq:fs}
    f_{\rm s} = K(z,T,N_{\rm H}) \dfrac{E(z)^3}{(1+z)^4} & \Big( \dfrac{k_{\rm B} T}{\kev}\Big) \\
    & {\rm photons}\, {\rm Ms}^{-1}\, {\cm}^{-2}\, (0.984\, {\rm arcsec})^{-2}, \nonumber
\end{align}
where $K(z,T,N_{\rm H})$ is the K-correction calculated with the APEC model as done in Section~\ref{sec:method-XrayMock} and $E(z) \equiv H(z)/H_{\rm 0}$. The scaling reduces the redshift and halo mass dependence of the surface brightness in observational samples. Assuming the self-similarity of relaxed clusters, it becomes possible to approximately identify corresponding regions of clusters with different masses and redshifts, without explicitly assuming the angular diameter distance to each or a prescription for estimating some scale radius. The isophotes of the images will be determined based on flux levels (in unit of $f_{\rm s}$) $S_{\rm j} = N_{\rm j} f_{\rm s}$, where we set the number of isophotes to three so $j=0,1,2,3$. $N_{\rm j}$ will be uniformly spaced in the logarithm, and the minimum and maximum levels ($N_{\rm 0}$ and $N_{\rm 3}$) will be tuned such that the radii of the isophotes roughly match the radial range of interest (see Section~\ref{sec:results-sb} for the tuning), $10-20\%$ $R_{\rm 200}$. After an adaptive smoothing of the original flat-fielded image, the code identifies pixels in isophotes based on pre-defined surface brightness levels $S_{\rm j}$. Then, an elliptical shape is fit to each of these isophotes, minimizing the sum of absolute distances from the ellipse to each pixel in the isophote along the line passing through the pixel and the ellipse center. The semi-major axis, center coordinates, position angle and ellipticity of each isophote are obtained. The uncertainties of the measured morphological parameters can be derived by performing the steps above on bootstrap realizations of each observation. Since the typical uncertainty in ellipticity is about two orders of magnitudes smaller than the halo-to-halo variation, in general we ignore it in the following analysis.

\section{Results}
\label{sec:results}

\begin{figure}
    \centering
    \includegraphics[width=0.49\textwidth]{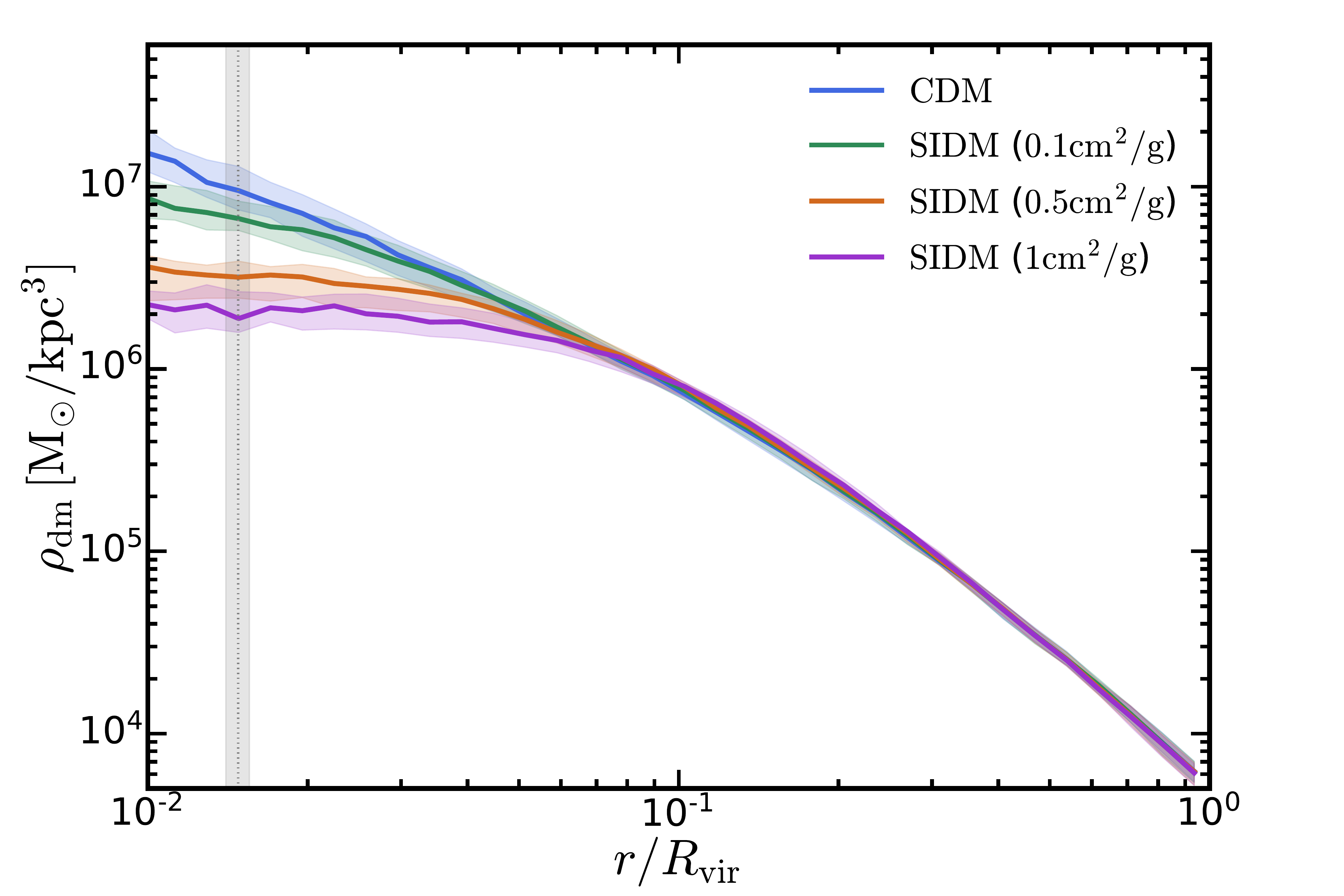}
    \caption{Dark matter mass density profiles of the simulated clusters. For each dark matter model, we show the median and $1\sigma$ dispersion of the density profiles of the simulated clusters. The grey dotted line with a shaded region indicates the conservative estimation of the convergence radii of dark matter properties with its error. SIDM haloes develop thermalized cores with flat central density profiles, in contrast to the cuspy central profile in CDM. The core size increases with greater self-interaction cross-sections. These differences exist outside the convergence radius, but eventually become negligible at the outskirts of the haloes ($\gtrsim 5\%\,R_{\rm 200}$).}
    \label{fig:dm_profile}
\end{figure}

\subsection{Density profile}
\label{sec:results-denpro}

In Figure~\ref{fig:dm_profile}, we show the dark matter mass density profiles of the simulated clusters. They are the average densities measured in uniformly spaced (in the logarithm) spherical shells. Both the median and $1\sigma$ dispersion of the density profiles are presented. The radius of convergence of dark matter properties is calculated using Equation~\ref{eq:power} in Section~\ref{sec:sim} and listed in Table~\ref{tab:sims}. For each cluster, we choose the maximum convergence radius from all four dark matter models as a conservative estimate. The median and $1\sigma$ dispersion of the convergence radii of all simulated clusters are shown with the vertical dashed line and the shaded region. Unlike the cuspy central profile in CDM, SIDM haloes develop flat and thermalized cores, with increasing core sizes with higher self-interaction cross-sections. Compared to the CDM case, the central dark matter density in the SIDM model with $(\sigma/m)=1\cpm$ is about five times (circa $0.7\,{\rm dex}$) lower at $r \sim 2\%\,R_{\rm 200}$. Even for the SIDM model with the lowest cross-section in the suite, $0.1\cpm$, the profile is distinguishable from the CDM case at the $2\sigma$ level outside the convergence radius. However, all the differences eventually diminish at the outskirts of the clusters, at larger than about $5\%\,R_{\rm 200}$. Although the discrepancy between SIDM and CDM predictions is significant at halo centers, contamination from gas cooling, star formation and feedback effects in the central galaxies is expected to be important in those regions. These factors will be discussed in detail in Section~\ref{sec:discussion}.

\begin{figure}
    \centering
    \includegraphics[width=0.49\textwidth]{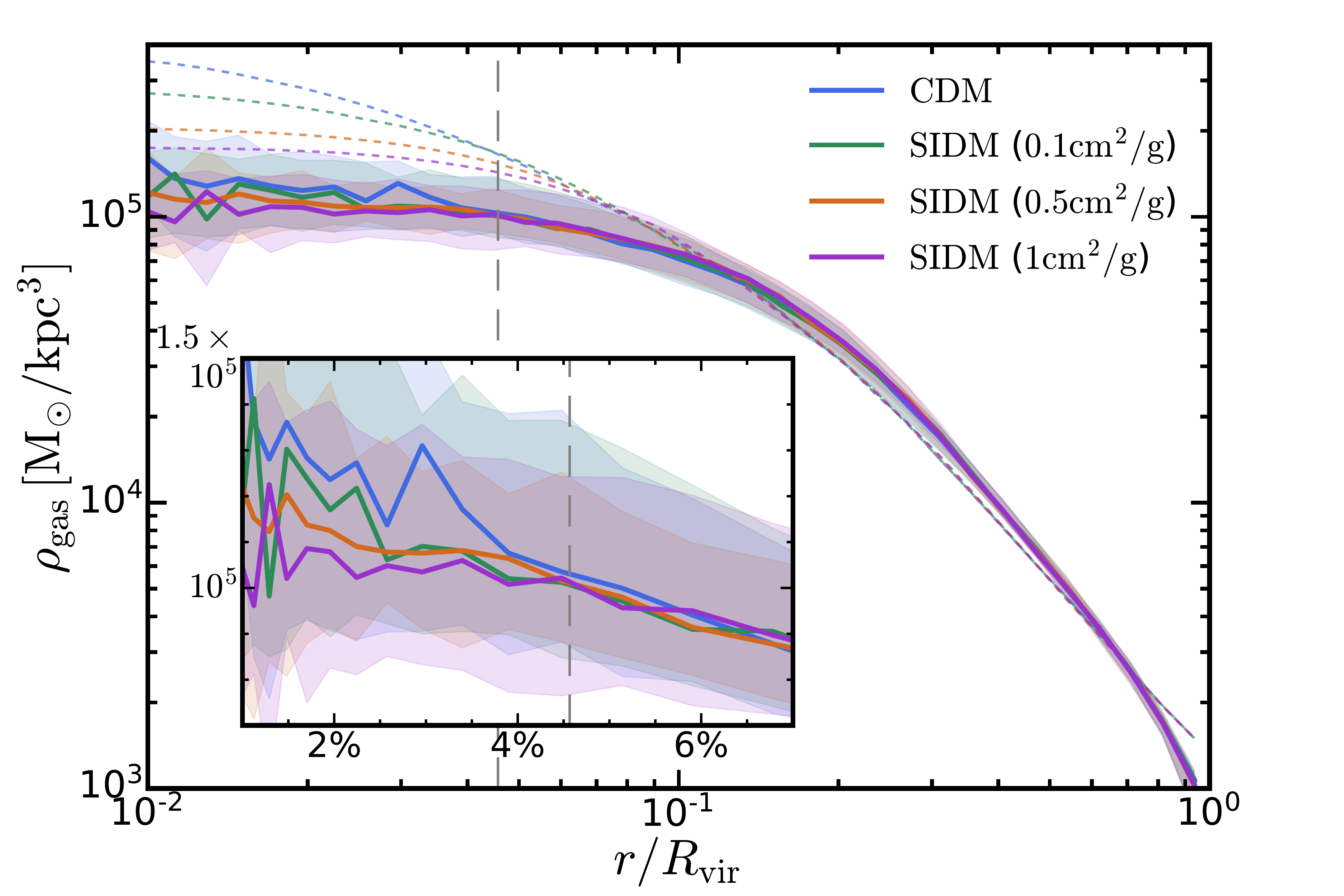}
    \caption{Gas mass density profiles of the simulated clusters. The labelling is the same as in Figure~\ref{fig:dm_profile}. The convergence radius for hydrodynamical properties of the gas is ambiguous, so we choose $16$ times the hydro spatial resolution $h_{\rm b}$ as a reference, indicated with the grey dashed line (see Figure~\ref{fig:gas_he} and the discussion at the end of Section~\ref{sec:results-denpro} for the convergence criterion). The colored short dashed lines show gas density profiles inferred from the gravitational potential of the gas, assuming that the intracluster gas is isothermal and in hydrostatic equilibrium. A zoom-in subplot is included to compare density profiles at the center. Unlike dark matter, the gas density profiles show little difference between dark matter models. The central densities are also lower than expected from the hydrostatic equilibrium predictions.}
    \label{fig:gas_profile}
\end{figure}

\begin{figure}
    \centering
    \includegraphics[width=0.49\textwidth]{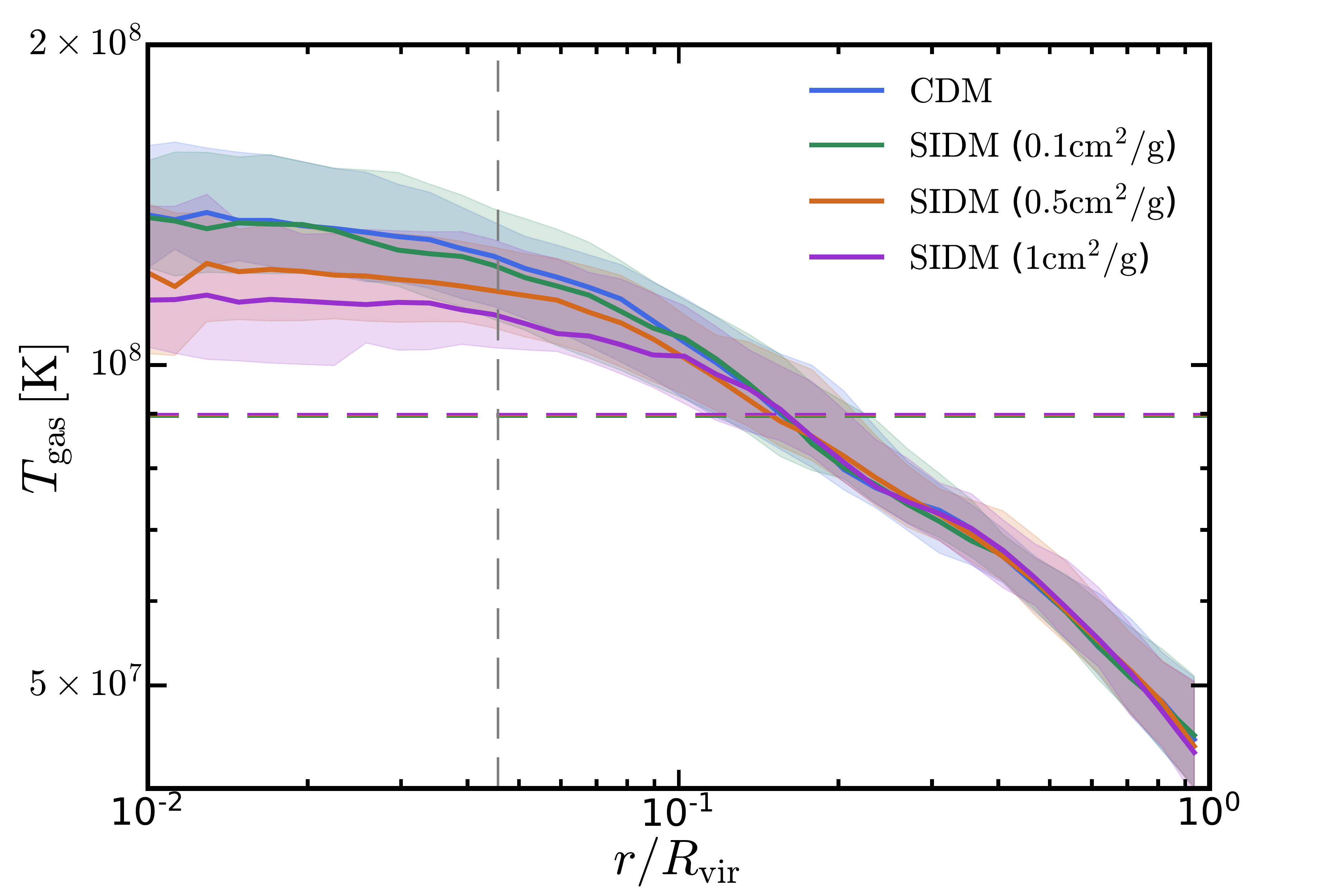}
    \includegraphics[width=0.49\textwidth]{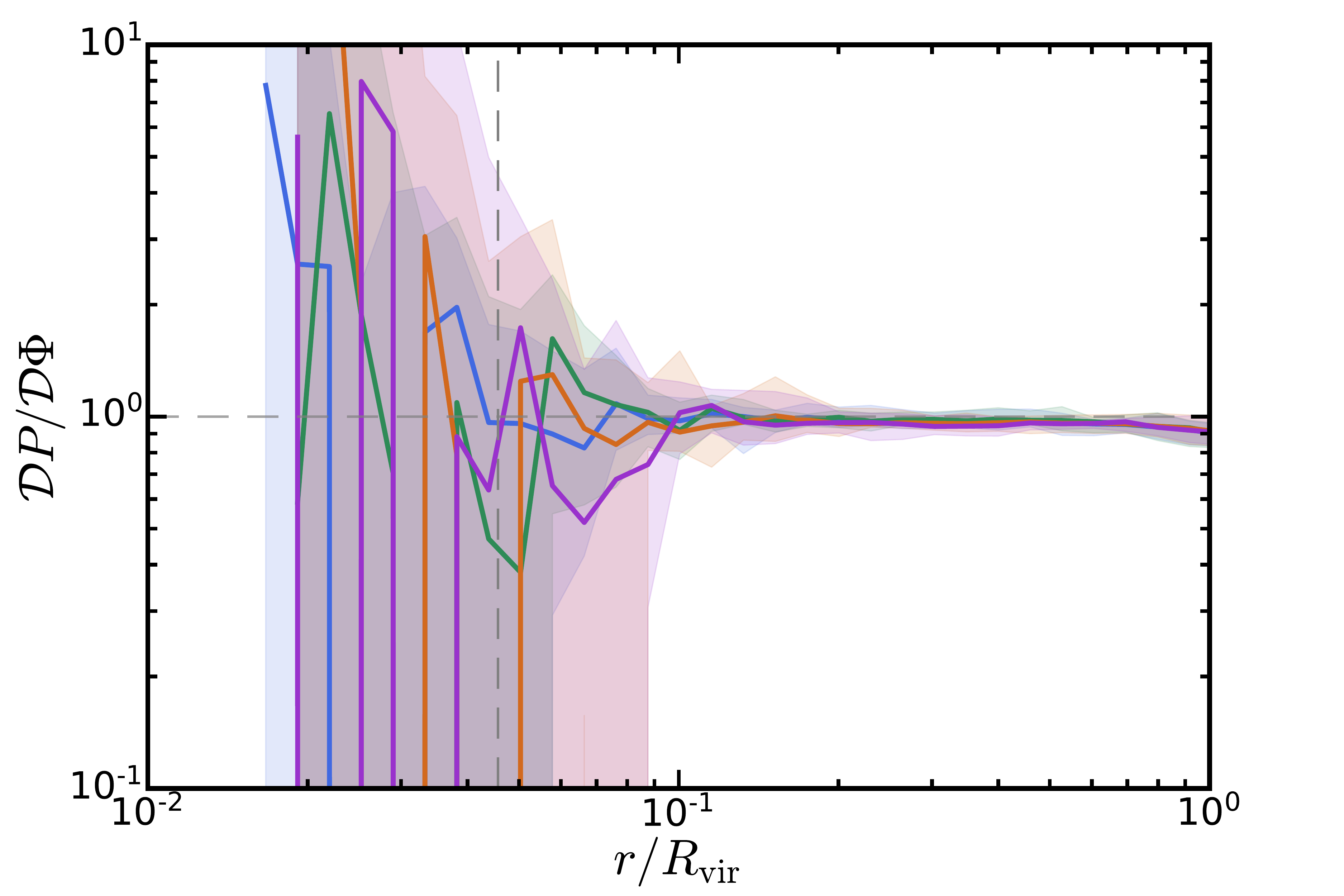}
    \caption{ {\it Top:} Gas temperature profiles of the simulated clusters. The labelling is the same as in Figure~\ref{fig:gas_profile}. The vertical dashed line shows an estimate of the convergence radius for hydrodynamical properties. The horizontal dashed lines indicate the median virial temperatures of the haloes in each dark matter model. Gas temperatures rise monotonically towards halo centers with the central temperature being slightly lower in SIDM models. {\it Bottom:} Thermal pressure gradient versus gravitational potential gradient as a function of radius. The thermal pressure support balances the gravitational attraction at $r\gtrsim 10\% \, R_{\rm 200}$, indicating that the hydrostatic equilibrium is perfectly respected. At small radii, the dispersion in $\mathcal{D}P/\mathcal{D}\Phi$ gradually becomes larger. The convergence radius for hydrodynamical properties is estimated as $16$ times the hydro spatial resolution $h_{\rm b}$ and is indicated with the grey vertical dashed line. Within the convergence radius, the median value of $\mathcal{D}P/\mathcal{D}\Phi$ shows order of magnitude fluctuations.}
    \label{fig:gas_he}
\end{figure}

We apply the same analysis to the intracluster gas in the simulations. In Figure~\ref{fig:gas_profile}, we show the gas mass density profiles of the simulated clusters. Assuming the intracluster gas is in hydrostatic equilibrium, the gas should distribute in a way that the thermal pressure balances the gravitational attraction (neglecting non-thermal pressure from, e.g., turbulent gas motions, which are subdominant in massive relaxed clusters, \citealt{Lau2009,Vazza2011,Nelson2014}). If we further assume that the gas is isothermal, the gas density is simply related to the gravitational potential, $\Phi$, as
\begin{equation}
    \dfrac{\rho_{\rm gas}(r)}{\rho_{\rm gas}(0)} = \exp{ \left[ - \dfrac{\mu m_{\rm p} \Phi(0)}{k_{\rm B}T}\big( \Phi(r)/\Phi(0)-1\big) \right] },
\end{equation}
where the isothermal temperature $T$ can be approximated as the virial temperature of the halo. In Figure~\ref{fig:gas_profile}, the profiles determined from the potential are shown in short dashed lines for reference. For both the gas mass density profile and the equilibrium-modelled gas density profile, the difference between different dark matter models is small, as opposed to the distinct signature of SIDM in the dark matter density profile. Part of the reason is that the gravitational potential is less sensitive to the dark matter density differences at small radii, thus the equilibrium-modelled gas density profiles are also less sensitive to SIDM physics. However, compared to the equilibrium-modelled ones, the gas mass density profiles are systematically lower at cluster centers and the SIDM related differences are also smaller. This is likely related to a deviation from hydrostatic or thermal equilibrium, which we will investigate in the following.

\begin{figure*}
    \centering
    \includegraphics[width=0.49\textwidth]{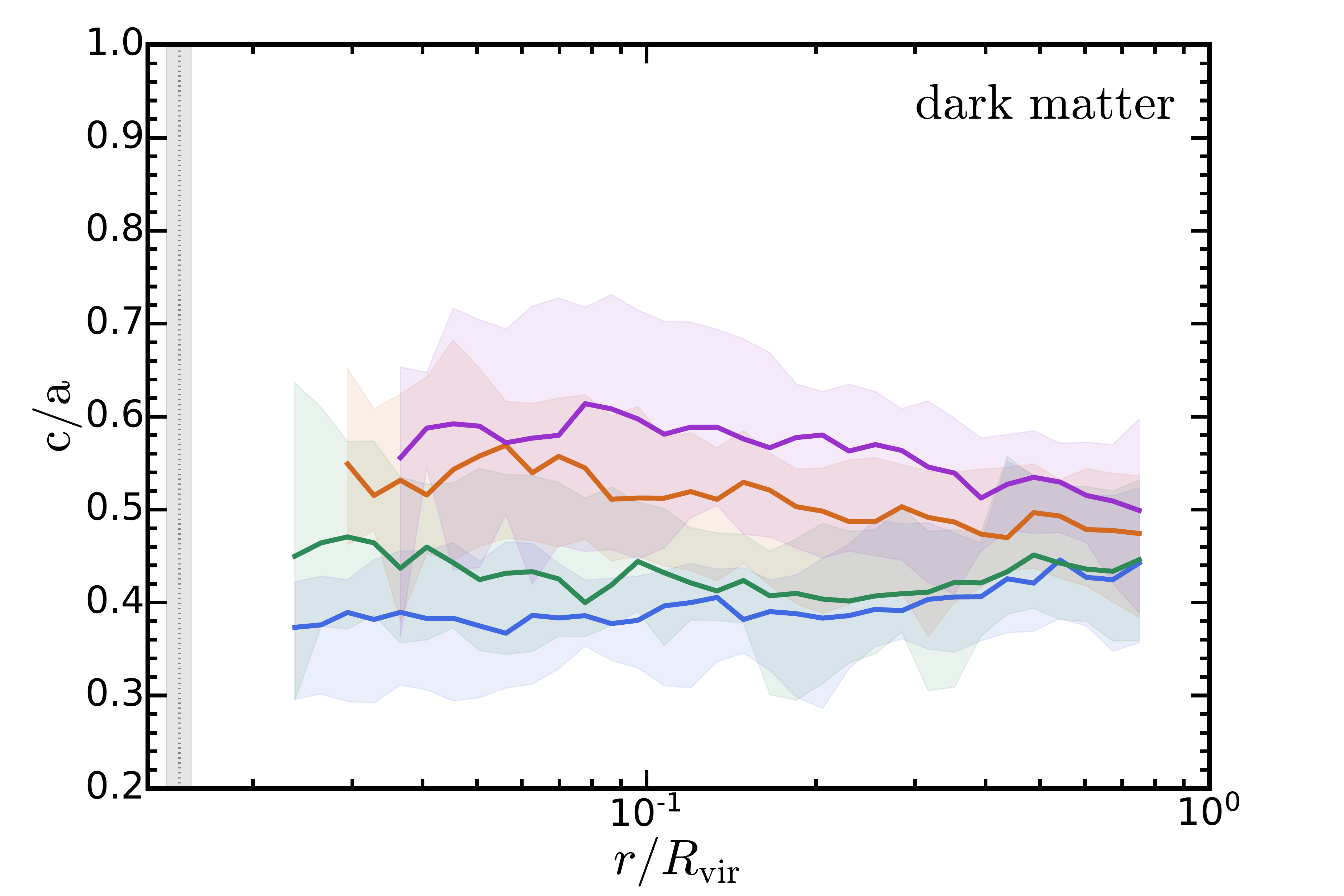}
    \includegraphics[width=0.49\textwidth]{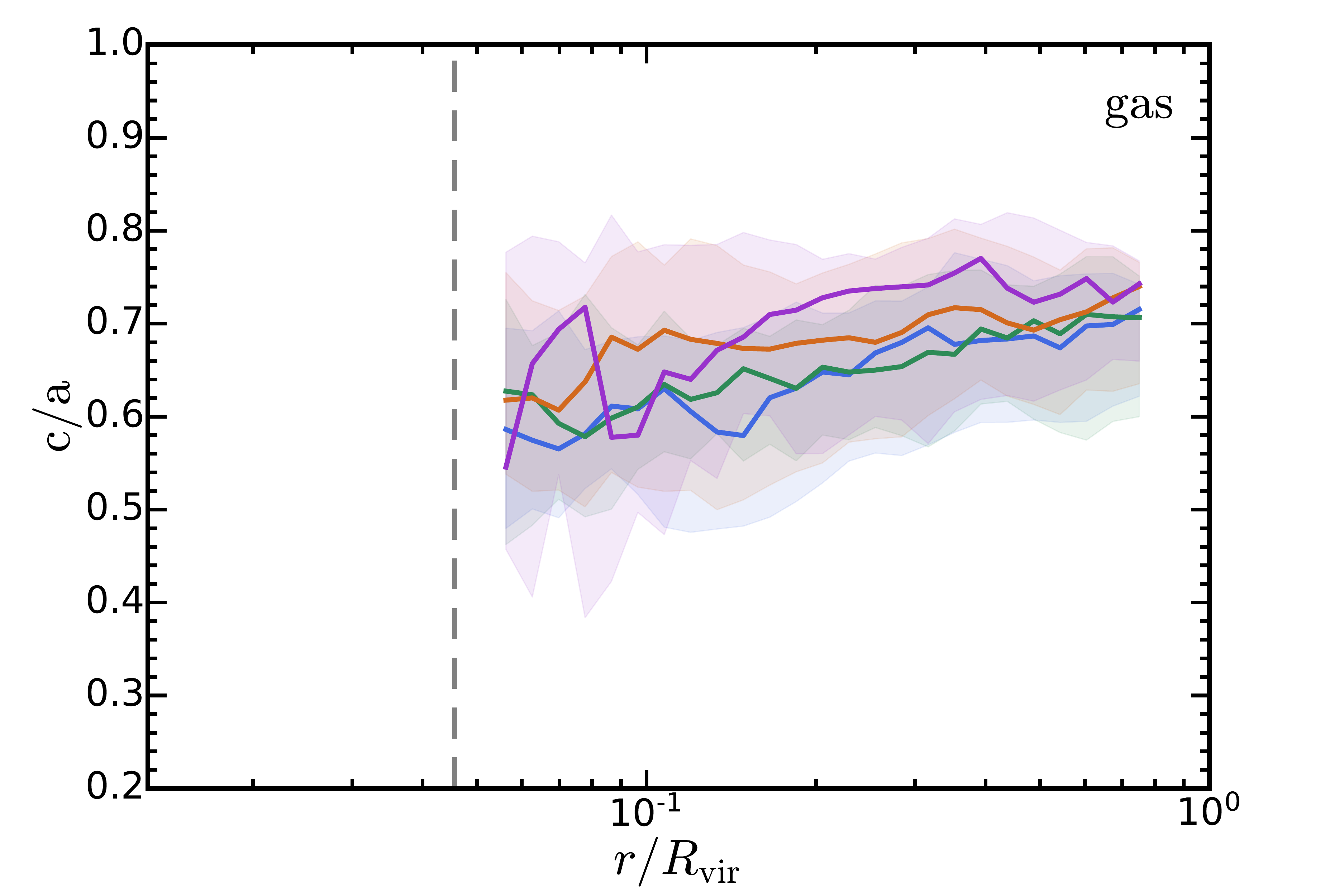}
    \includegraphics[width=0.49\textwidth]{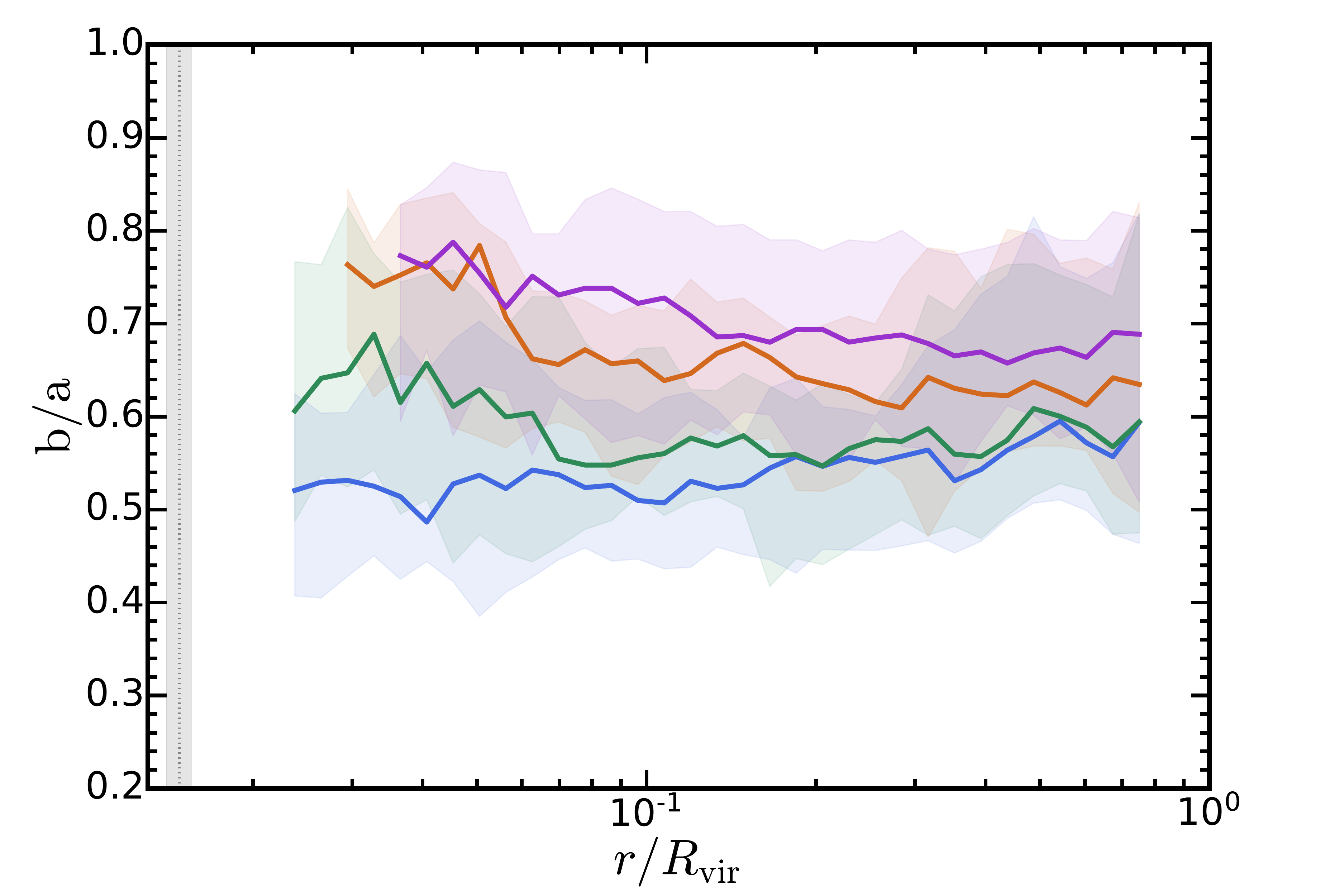}
    \includegraphics[width=0.49\textwidth]{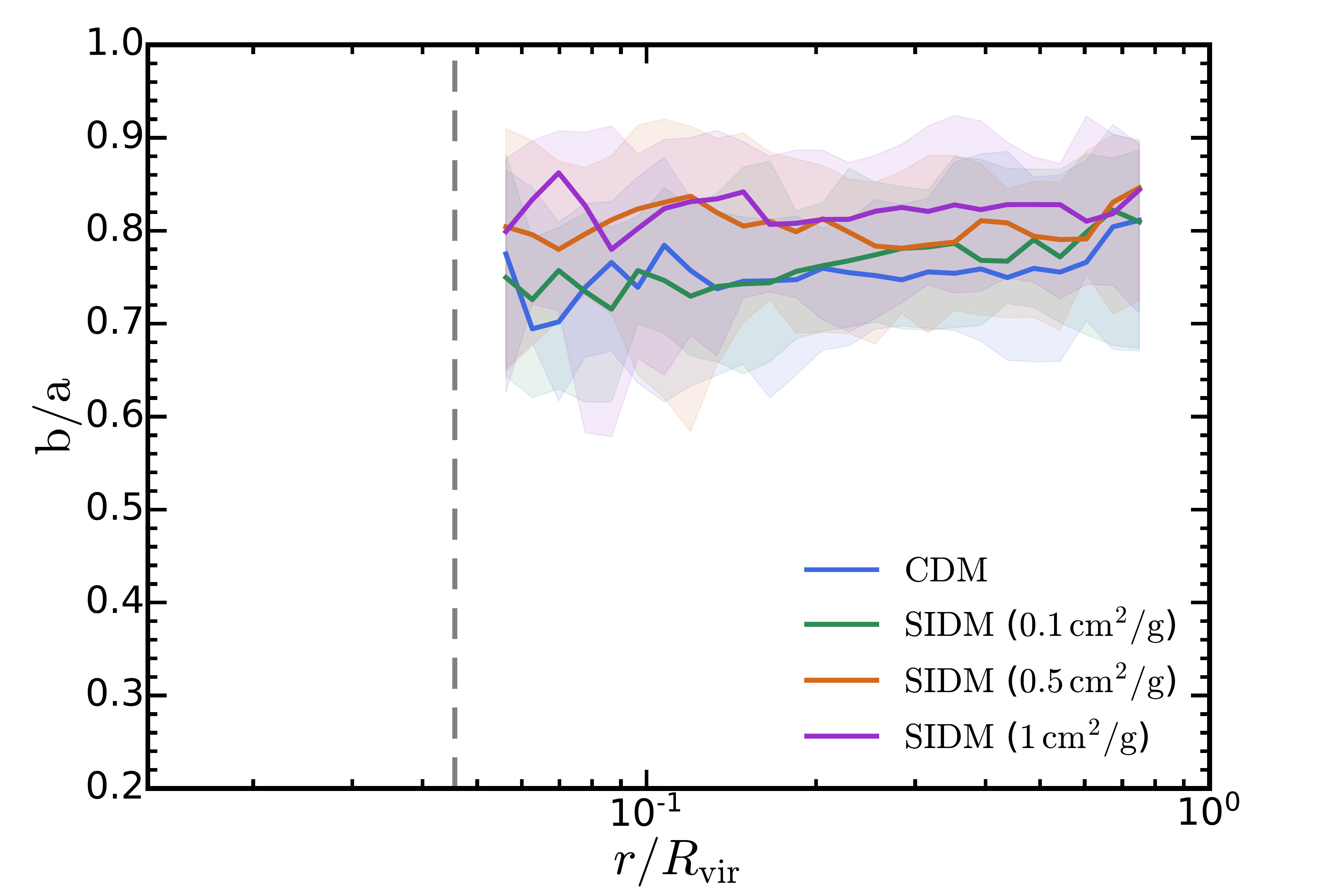}
    \caption{Three-dimensional axial ratios of the dark matter (gas) distribution of the simulated clusters. The left (right) column shows the axial ratios for dark matter (gas). The top row shows the minor-to-major axial ratio, $c/a$, and the bottom row shows the intermediate-to-major axial ratio, $b/a$. The shaded vertical regions on the left and dashed lines on the right indicate the convergence radii for dark matter and gas properties, respectively. Considerable differences between SIDM and CDM show up in the dark matter shape out to large radii, while the distinct signature of SIDM in the gas shape is much weaker. Meanwhile, the gas distribution is systematically rounder than the dark matter one, as a consequence of the X-ray emitting gas tracing more directly the isopotential surface of the matter distribution rather than the mass distribution.}
    \label{fig:3dshape}
\end{figure*}

Assuming spherical symmetry and neglecting non-thermal pressure support, the hydrostatic equilibrium implies
\begin{align}
    \dfrac{\partial \Phi(r)}{\partial r} & = - \dfrac{1}{\rho_{\rm gas}(r)}  \dfrac{\partial P(r)}{\partial r} \nonumber \\
    & = - \dfrac{k_{\rm B} T(r)}{r \mu m_{\rm p}}  \left(\dfrac{\partial \ln{\rho_{\rm gas}(r)}}{\partial \ln{r}} + \dfrac{\partial \ln{T(r)}}{\partial \ln{r}} \right),
    \label{eq:he}
\end{align}
which simply represents that the gravitational attraction is balanced by the thermal pressure induced by either the density or temperature gradient. We denote the left hand side of the equation as ``$\mathcal{D}\Phi$'' and the right hand side as ``$\mathcal{D}P$''. In perfect hydrostatic equilibrium, we expect $\mathcal{D}P = \mathcal{D}\Phi$. In the top panel of Figure~\ref{fig:gas_he}, we show the gas temperature profiles of the simulated clusters. With the absence of cooling processes, the gas temperature rises monotonically towards the cluster center, in line with the picture that the infalling gas is heated by strong accretion shocks. As described in Equation~\ref{eq:he}, the temperature gradient revealed here contributes to the thermal pressure support of gas and, as a result, the gas mass density profile rises slower than the isothermal profile towards the cluster center. Apart from this, SIDM models give slightly lower gas temperatures at $r\lesssim 0.1\,R_{\rm 200}$ and the temperature gradients are also smaller, which makes the differences between gas density profiles in SIDM and CDM even smaller. In the bottom panel of Figure~\ref{fig:gas_he}, we show $\mathcal{D}P/\mathcal{D}\Phi$ as a function of radius. To obtain $\mathcal{D}P$, the pressure and temperature gradients are evaluated between adjacent spherical shells. The hydrostatic equilibrium is perfectly respected at $r\gtrsim 10\% \, R_{\rm 200}$. The dispersion of $\mathcal{D}P/\mathcal{D}\Phi$ gradually becomes larger at smaller radii and SIDM models in general show greater dispersion. This dispersion is likely caused by the limited statistics of gas cells. For reference, the grey vertical dashed line indicates $16$ times the hydro spatial resolution $h_{\rm b}$ (the equivalent size of gas cells), which roughly corresponds to $8$ times the spatial spacing of gas cells. Within this reference radius, the median $\mathcal{D}P/\mathcal{D}\Phi$ in all dark matter models starts to deviate significantly from unity and exhibits order of magnitude oscillations. Therefore, we choose this radius as the convergence radius of hydrodynamical properties of the gas. This radius is also plotted in Figure~\ref{fig:gas_profile} and the top panel of Figure~\ref{fig:gas_he} as reference for convergence.

\subsection{Shapes of dark matter and gas distributions}
\label{sec:results-shape}

\citet{Brinckmann2018} found that the three-dimensional shape of dark matter haloes is quite sensitive to SIDM physics. However, as demonstrated in the previous section, gas properties in general are much less sensitive to SIDM physics compared to dark matter. Therefore, it is important to check whether the shape changes in the dark matter distributions are reflected at the same level in the shape of the gas distribution.

To study the shapes of dark matter or gas distributions in simulated clusters, we adopt the code developed by \citet{Brinckmann2018} based on the methodology in \citet{Zemp2011}. The code determines the orientation and magnitude of the principal axes of a distribution of particles by computing the eigenvectors and eigenvalues of the shape tensor, defined as
\begin{equation}
    {\bf S} \equiv \dfrac{\int_{V} \rho\, {\bf r}\, {\bf r}^{T} {\rm d}V}{\int_{V} \rho\, {\rm d}V},
\end{equation}
where $\rho$ is the density, ${\bf r}$ is the position vector relative to the halo center and ${\bf r}^{T}$ is the transpose of it. The discrete form of the shape tensor is defined as
\begin{equation}
    S_{\rm ij} \equiv \dfrac{ \sum_{\rm k} m_{\rm k}\,  r_{\rm k}^{\rm i}\, r_{\rm k}^{\rm j} }{\sum_{\rm k} m_{\rm k}},
\end{equation}
where $m_{\rm k}$ is the mass of the {\it k}th particle and $r_{\rm k}^{\rm i}$ is the {\it i}th component of the position vector of the {\it k}th particle. In our analysis, we divide each halo into a number of ellipsoidal shells. The shells are initialized as spherical and are adaptively merged or split based on the particle number in each shell (adjacent shells with less than $2000$ particles are merged and shells containing $50000$ or more particles are split). The code computes the eigenvectors and eigenvalues of the shape tensor of the particles within each shell, until convergence is achieved (when the axial ratios of both the minor and intermediate axes to the major axis deviate by less than one percent over the last ten iterations). For each iteration, the volume of the ellipsoidal shell will deform according to the axes determined in the previous loop, with particles being added or removed from the shell accordingly, while keeping the length of the major axis invariant. After convergence is reached, we document the minor-to-major axial ratio $c/a$ and the intermediate-to-major axial ratio $b/a$ for each shell, and compute the effective radius of the shell as $r_{\rm eff} = a \sqrt{[(c/a)^2+(b/a)^2+1]/3}$, similarly to what we have done in the isophote analysis (see Section~\ref{sec:method-2dshape}). We apply the method described above to both, the dark matter particles and gas cells in our simulated clusters.

In Figure~\ref{fig:3dshape}, we show the three-dimensional axial ratios of the dark matter and gas distributions of the simulated clusters. For each model, we again present the median and $1\sigma$ dispersion of the axial ratios. Similar to what was found in \citet{Brinckmann2018}, we see that $c/a$ for dark matter in the SIDM-c1 model can deviate from the CDM case at $2\sigma$ level out to about $0.2\,R_{\rm 200}$. Note that at a similar radius the density profiles in SIDM and CDM are already indistinguishable, as shown in Figure~\ref{fig:dm_profile}. Even for the SIDM-c0.1 model, $c/a$ for dark matter is distinguishable at about $1\sigma$ level out to $0.1\,R_{\rm 200}$. These findings are consistent with other cosmological simulations of cluster-mass haloes in SIDM~\citep[e.g.,][]{Peter2013,Robertson2019}. On the contrary, gas shape differences between SIDM and CDM become much weaker and the shape profiles systematically rounder than for dark matter. For example, at $r \sim 0.2\,R_{\rm 200}$, we see that $c/a$ in the SIDM-c1 model deviates from the CDM prediction at only about $1\sigma$ level. In hydrostatic equilibrium, the isodensity (and isotemperature) surface of the gas distribution should trace the isopotential surface of the matter distribution, also known as the {\it X-ray shape theorem} \citep{Buote1994}; $\mathop{\nabla} \rho_{\rm gas} \times \mathop{\nabla} \Phi = 0$. Since the isopotential surfaces are typically rounder than the source matter distribution \citep[e.g.,][]{Binney2008,Morandi2010,Limousin2013}, the shape of the gas distribution is rounder than dark matter as a consequence. In \citet{Robertson2019}, it was found that the stellar and gas distributions in SIDM with $(\sigma/m)\lesssim1\cpm$ and CDM show almost no difference in shape. However, in our results we still find some residual differences between SIDM and CDM that could be tested statistically with large samples of simulated and observed galaxy clusters. 

\subsection{X-ray surface brightness profile}
\label{sec:results-sb}

\begin{figure}
    \centering
    \includegraphics[width=0.49\textwidth]{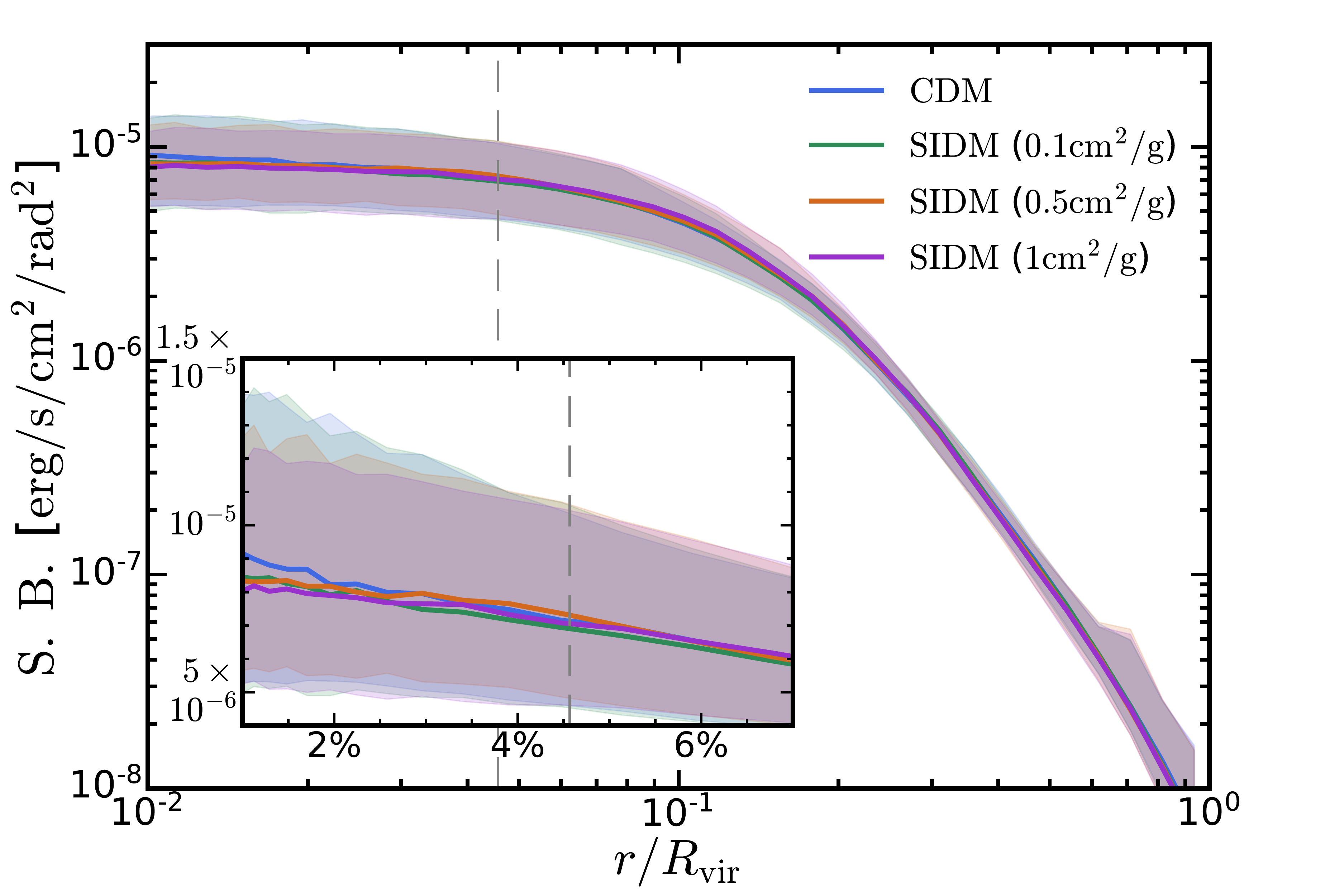}
    \caption{Soft X-ray surface brightness profiles of the simulated clusters in different dark matter models. For each model, we show the median and $1\sigma$ dispersion of the surface brightness profiles. A zoom-in subplot of the central surface brightness profiles is included. The hydro convergence radius is shown with the grey vertical dashed line. The surface brightness profile is basically insensitive to dark matter physics, due to a combination of projection effects and the weak response of the intracluster gas distribution to SIDM physics.}
    \label{fig:sbmodel}
\end{figure}

\begin{figure}
    \centering
    \includegraphics[width=0.49\textwidth]{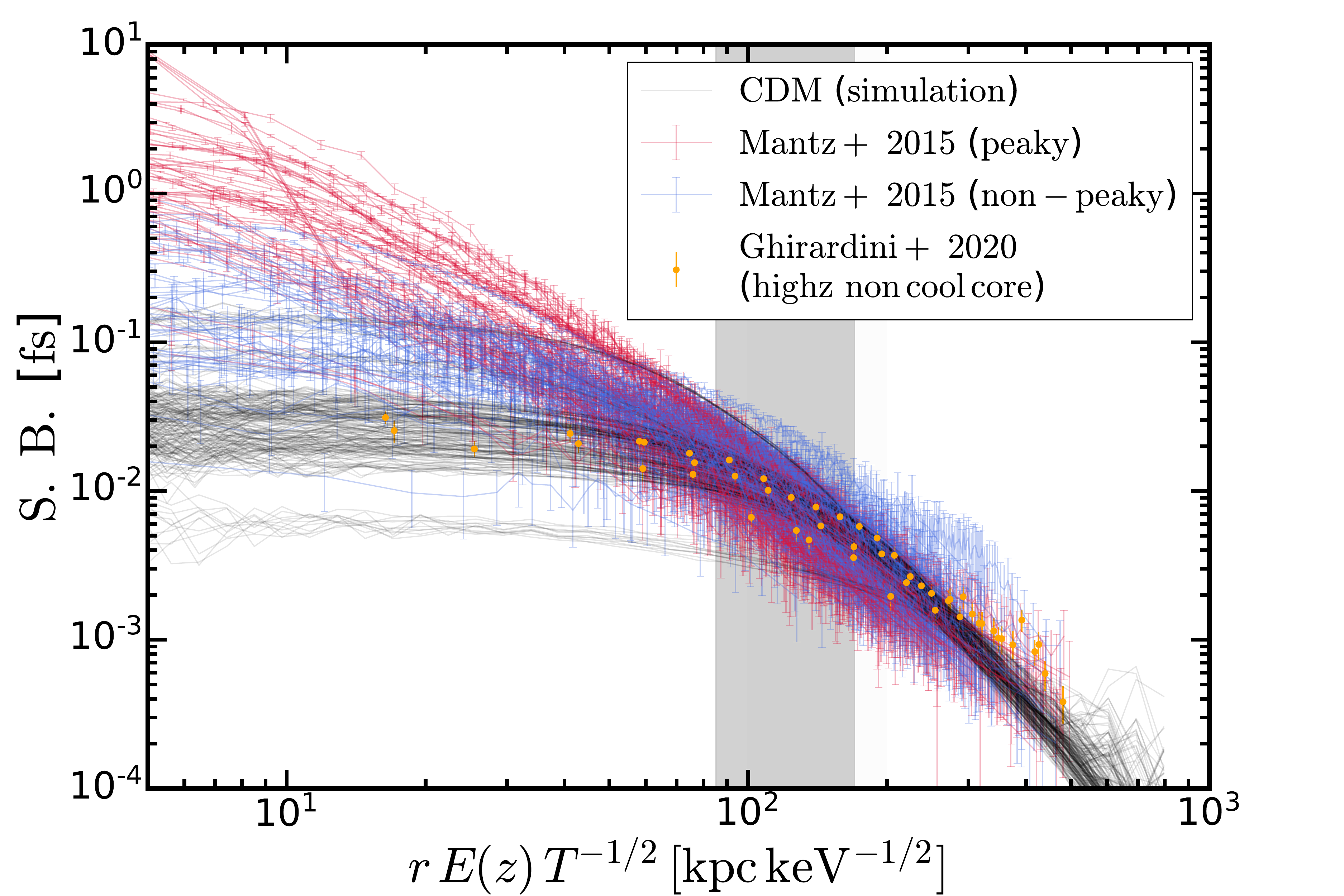}
    \includegraphics[width=0.49\textwidth]{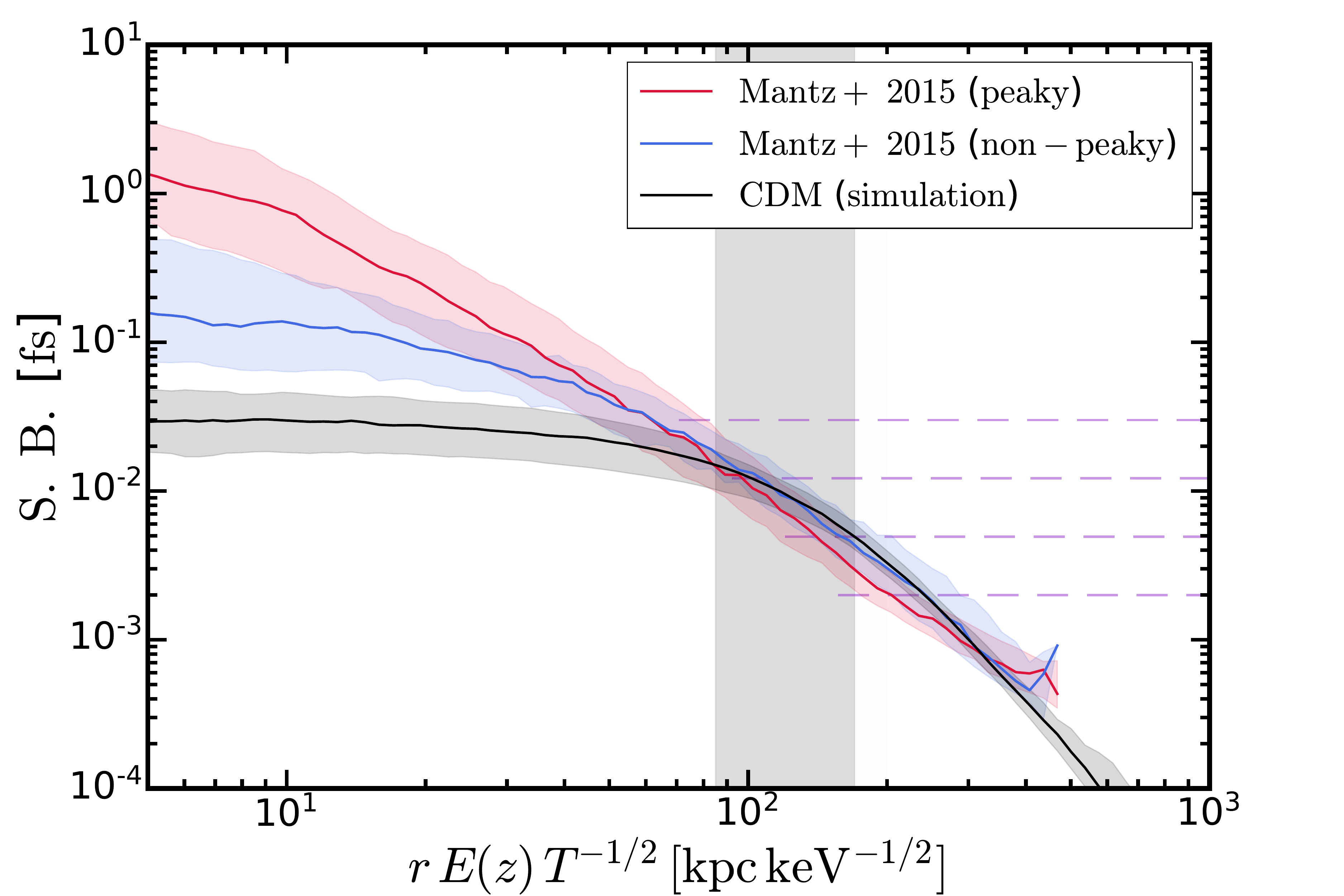}
    \caption{Comparisons of the soft X-ray surface brightness profiles of the simulated and observed clusters. {\it Top}: Surface brightness profiles (with observational error bars) of individual clusters. The observed clusters are grouped as the low-redshift ``peaky'' and ``non-peaky'' clusters \citep{Mantz2015}, and the three SPT-selected clusters at $z\gtrsim 1.2$~\citep{Ghirardini2021} which do not exhibit cool-cores. The surface brightness profiles are normalized with the units defined in Equation~\ref{eq:fs}. The shaded region indicates the radial range of interest, $0.1\operatorname{-}0.2\,R_{\rm 200}$. The cuspy central profiles of the low-redshift observed clusters, in particular the ``peaky'' sample, are not present in the simulated clusters with the absence of cooling processes. The high-redshift SPT-selected clusters appear to agree better with the simulations, due to the different thermodynamical properties compared to the low-redshift clusters. At large radii, including the radial range of interest for shape measurements, we find reasonable agreement in terms of normalization and slope between the simulated and observed profiles. {\it Bottom}: Median and $1\sigma$ dispersion of each group of surface brightness profiles. It is clear that the simulated clusters agree better with the ``non-peaky'' sample at the outskirts of the clusters. Based on the surface brightness in the radial range of interest ($0.1\operatorname{-}0.2\,R_{\rm 200}$, indicated with the shaded region), we pick the flux levels of the isophotes for morphology analysis. They are marked by the purple dashed lines.}
    \label{fig:sbobs}
\end{figure}

The next question to answer is how the differences in the three-dimensional shape of gas are reflected in the two-dimensional shape of X-ray isophotes. To answer this, we first create mock X-ray images for the simulated clusters following the procedure described in Section~\ref{sec:method-XrayMock} and measure the surface brightness profile. For each dark matter model and each halo, $12$ images are generated corresponding to $12$ sampled viewing angles. In Figure~\ref{fig:sbmodel}, we show the median and $1\sigma$ dispersion of the soft X-ray surface brightness profiles from the simulated clusters. Similar to what has been found for the gas density profiles (see Fig.~\ref{fig:gas_profile}), SIDM and CDM predictions are nearly indistinguishable. For a given spherical annulus, the projection effects will mix the gas emission at small and large three-dimensional radii, which makes the surface brightness profiles more cored than the density profiles at $r\lesssim 0.1\, R_{\rm 200}$ and further decreases the difference between SIDM and CDM. 

\begin{figure}
    \centering
    \includegraphics[width=0.49\textwidth]{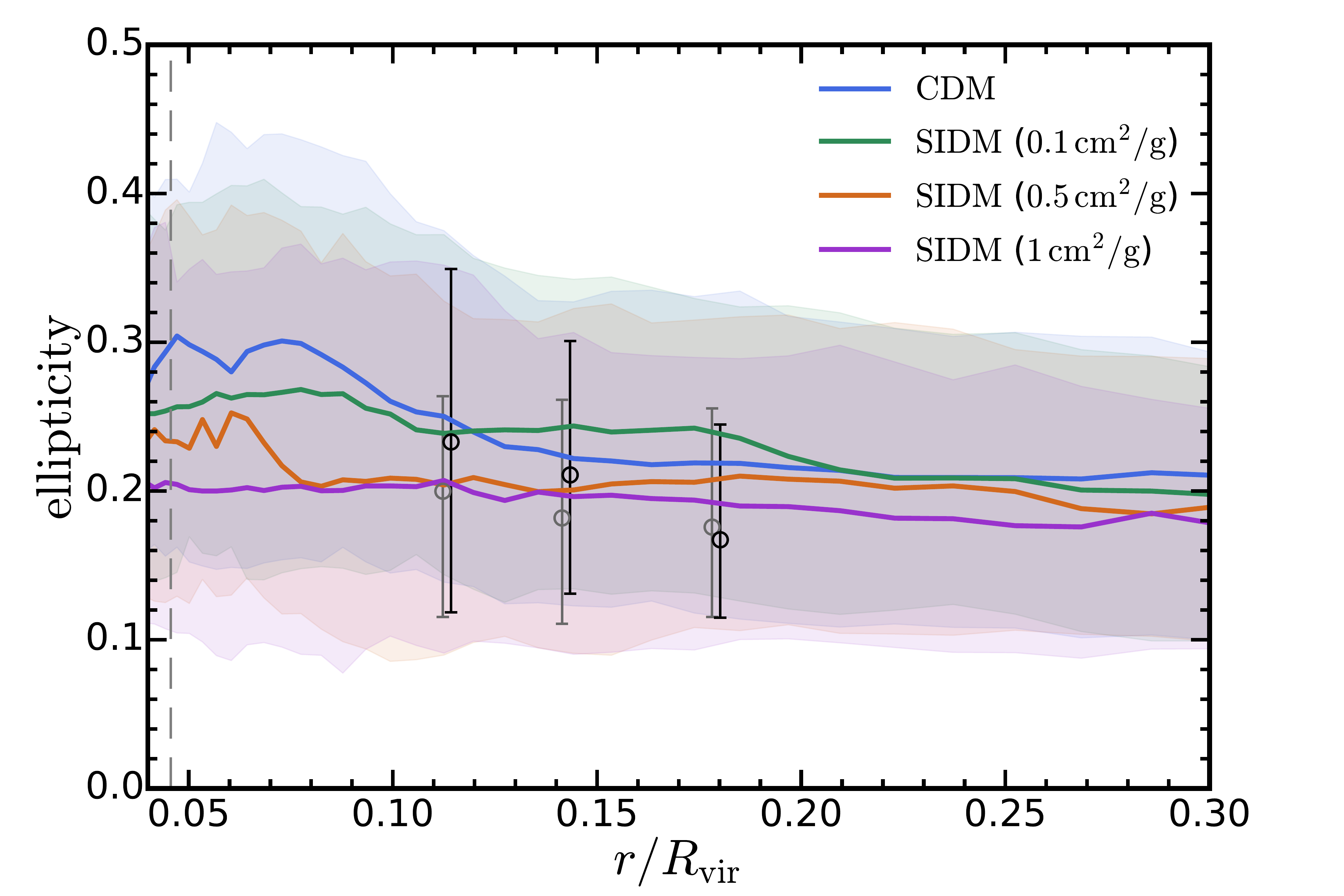}
    \caption{Ellipticity profiles of the simulated clusters compared to the observational results. We show the ellipticity of the isophotes as a function of the effective radius of the isophote. The median values and $1\sigma$ dispersions of the simulated samples are shown as solid lines and shaded regions. The results of the observed ``non-peaky'' (``peaky'') samples are shown by open black (grey) markers with error bars.  The vertical dashed line on the left indicates the hydro convergence radius. The SIDM-c0.5 and SIDM-c1 models predict lower ellipticities and agree better with the observational results. However, the signal is smeared by the large statistical uncertainties.}
    \label{fig:eps_profile}
\end{figure}

\begin{figure*}
    \centering
    \begin{minipage}{0.54\textwidth}
    \includegraphics[width=1.0\textwidth,trim={0.5 0 3cm 3cm},clip]{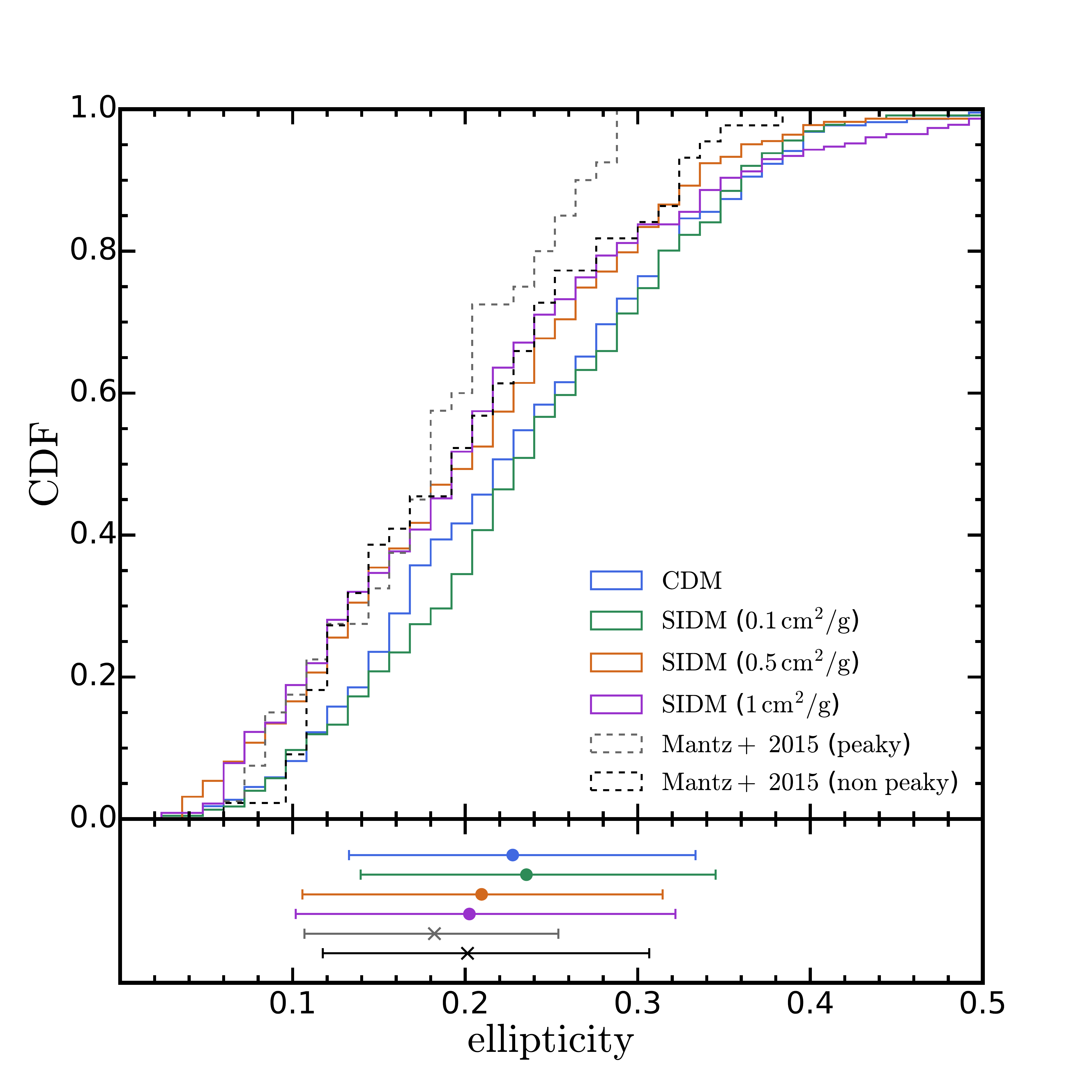}
    \end{minipage}
    \begin{minipage}{0.45\textwidth}
    \includegraphics[width=1.0\textwidth,trim={0.8cm 0 0 0cm},clip]{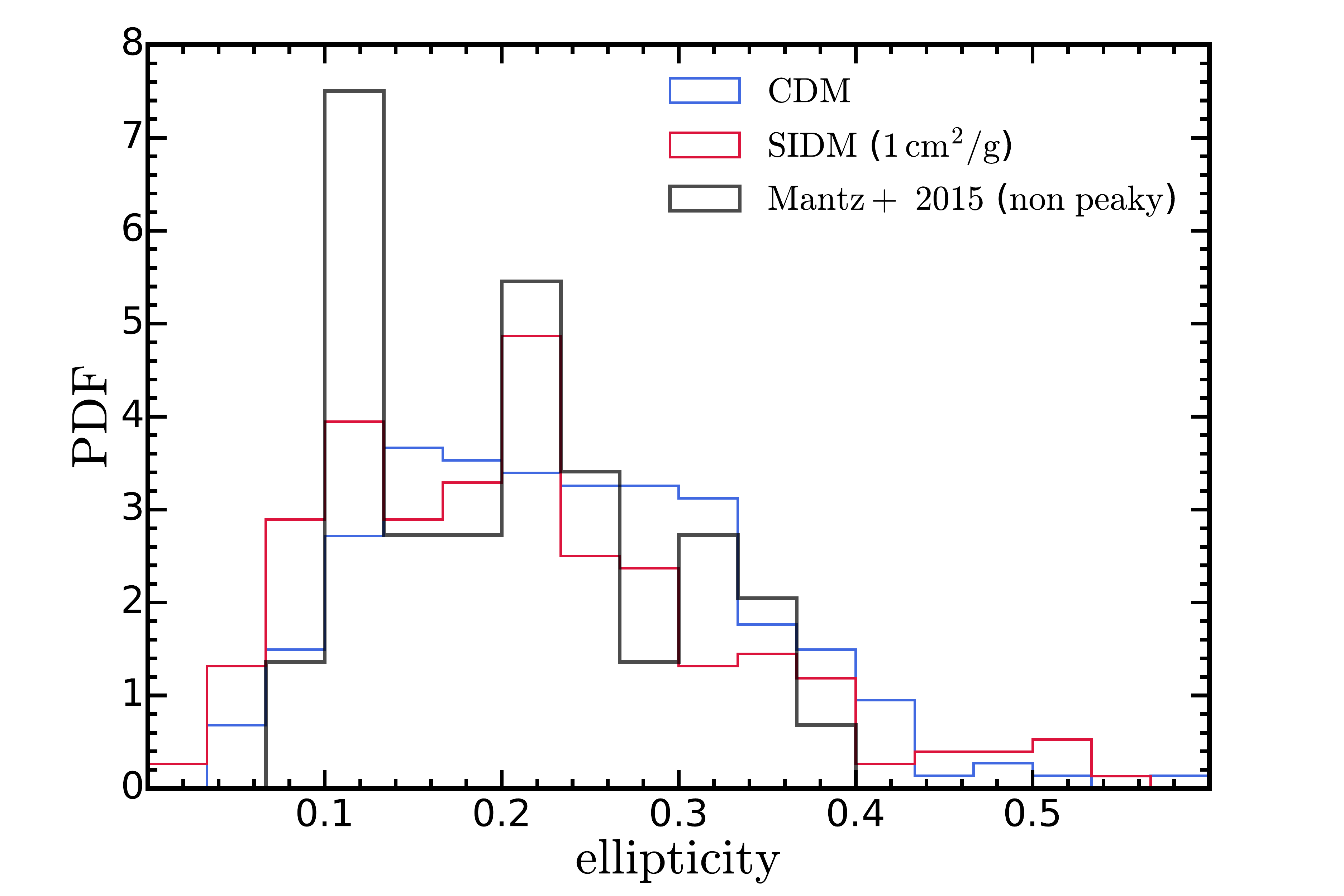}
    \includegraphics[width=1.0\textwidth,trim={0.8cm 0 0 0cm},clip]{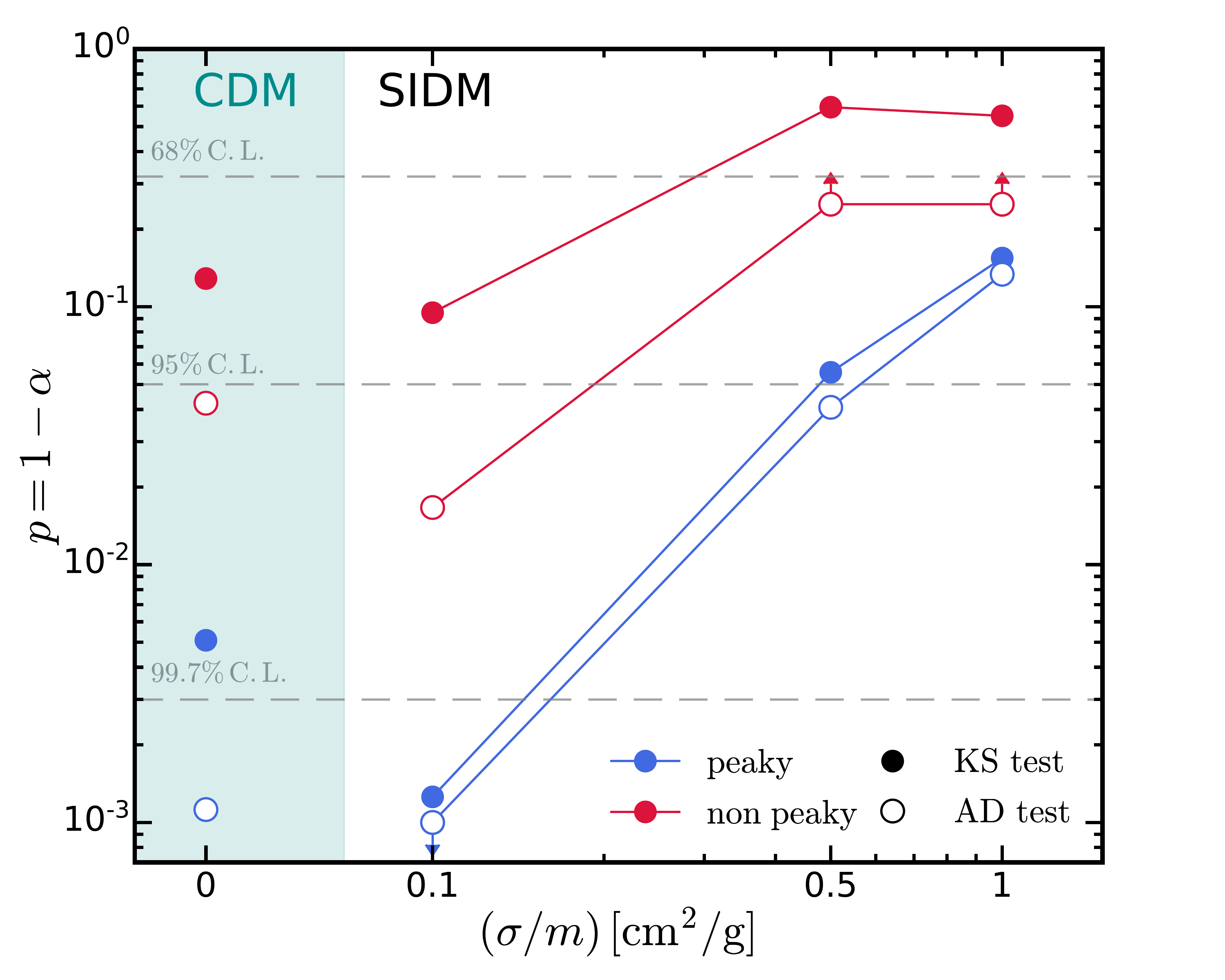}
    \end{minipage}
    
    \caption{{\it Left:} Cumulative distribution function of the ellipticities of the X-ray isophotes. The fit to the isophotes is performed at $0.1\operatorname{-}0.2\,R_{\rm 200}$ for each viewing angle of each simulated halo. In the lower subpanel, we show the median ellipticities and $1\sigma$ dispersions for the different dark matter models and the observed cluster samples. {\it Top right:} Probability distribution function of the ellipticities. For simplicity, we only show PDFs of the CDM and SIDM-c1 models compared to observations. {\it Bottom right:} $p$ value of the two-sample KS and AD tests. The tests are performed on the ellipticity distributions of observed and simulated clusters. The $p$ value is the likelihood that the two samples are drawn from the same underlying continuous distribution function. Compared to the ``non-peaky'' sample, the KS and AD tests reject the CDM and SIDM-c0.1 models at about $90\%$ confidence level. }
    \label{fig:eps_cdf}
\end{figure*}

The soft X-ray surface brightness profiles can be directly compared to observational results. For the observed clusters introduced in Section~\ref{sec:obs}, the surface brightness profiles are measured using the {\sc Spa} code introduced in Section~\ref{sec:method-2dshape}. In the top panel of Figure~\ref{fig:sbobs}, we compare the surface brightness profiles of the simulated clusters with the observed ones, the latter of which are grouped as the ``peaky'' (red) and ``non-peaky'' (blue) samples. For clarity, we only show the results of the CDM simulations, since we have shown above that the surface brightness profiles are insensitive to SIDM physics. We show explicitly the profile for each observed cluster, with the observational uncertainties as error bars, along with the profile for each image of the simulated clusters (recalling that we have multiple possible projection angles for each simulated cluster). The annulii radii and the surface brightness are normalized following the convention in \citet{Mantz2015} to reduce potential redshift or temperature dependences. For reference, we show the profiles of three SPT-selected clusters at $z\gtrsim 1.2$ from \citet{Ghirardini2021}\footnote{Modelled surface brightness profiles convolved with the PSF matrix and then fitted to the raw data, considering the exposed area and time for each annulus as well as the background.}, which have distinct thermodynamical properties from the low-redshift clusters and do not exhibit cool cores. The clusters in the ``peaky'' observational sample have cuspy central profiles, in contrast with the cored profiles of our (adiabatic) simulated clusters. The ``non-peaky'' clusters are less cuspy, but the central surface brightness is still almost an order of magnitude higher than that of our simulated clusters. Since the simulations do not include gas cooling and physics of star formation and evolution, it is expected that the condensation of baryons at the center of clusters will be weaker for the simulated clusters compared to the observed ones. The shape of the surface brightness profile simply manifests the thermodynamical properties of the clusters. This statement is supported by the agreement of the simulation results with the non-cool-core clusters selected at high redshift. Despite the dissimilarity at small radii, the surface brightness profiles of all samples agree well with each other at large radii, including the radial range of interest for this work ($0.1\operatorname{-}0.2\,R_{\rm 200}$). In the bottom panel of Figure~\ref{fig:sbobs}, we condense the profiles shown on the upper panel to the median and $1\sigma$ dispersion of each sample. At $r\simeq 0.1\operatorname{-}0.2\,R_{\rm 200}$, the simulation results are in better agreement with the ``non-peaky'' observed sample. The ``peaky'' clusters have slightly lower surface brightness at the radius of interest, but the differences are small (less than about $0.1\,{\rm dex}$). Based on the surface brightness of simulated and observed clusters at $r\simeq 0.1\operatorname{-}0.2\,R_{\rm 200}$, we choose the flux levels for isophotes generation. Adopting the normalization convention in Section~\ref{sec:method-2dshape}, $N_{\rm 0}$ and $N_{\rm 3}$ are determined as $2\times 10^{-3}$ and $3\times 10^{-2}$, respectively. The flux levels bounding the three isophotes are marked as purple dashed lines in the figure.

\subsection{Ellipticity of the isophotes}
\label{sec:results-ellip}

Given the flux levels determined above, we use the {\sc Spa} code to select pixels for each isophote from the observed cluster images, and perform ellipse fitting to the isophotes as described in Section~\ref{sec:method-2dshape}. For the images generated from the simulated clusters we also perform ellipse fitting, as described in Section~\ref{sec:method-2dshape}. In Figure~\ref{fig:eps_profile}, we show the ellipticity of the isophotes as a function of the effective radius of the isophote (as defined in Section~\ref{sec:method-2dshape}). For the simulation results, the median and $1\sigma$ dispersion are shown for each dark matter model. For the observational results, the measured ellipticity of each isophote is shown and the radius, $r/R_{\rm 200}$, is determined from the comparison of surface brightness profiles in Figure~\ref{fig:sbobs}. Compared to the three-dimensional case, the two-dimensional shapes of the isophotes are much less sensitive to SIDM physics, primarily due to projection effects. First, a projected quantity (e.g. surface density, surface brightness) at a given projected radius $r_{\rm 2d}$ gets contribution from all three-dimensional radii at $r_{\rm 3d}>r_{\rm 2d}$. This ``mixing'' of information at different radii could mitigate signal strength. In addition, observed in different lines-of-sight, the same three-dimensional mass/luminosity distribution can appear to have different projected shapes, which acts as an additional source of noise. As shown in Figure~\ref{fig:eps_profile}, the SIDM-c1 and SIDM-c0.5 models are still distinguishable from CDM and the SIDM-c0.1 model, but the difference is smeared by large halo-to-halo variations and thus has low statistical significance. The ellipticities of the observed ``non-peaky'' clusters show a somewhat stronger radial dependence than the simulated clusters and the observed ``peaky'' clusters, in addition to exhibiting slightly larger ellipticities at smaller radii. Nevertheless, the results for the ``non-peaky'' sample are still more consistent with large cross-section SIDM-c1 and SIDM-c0.5 models than with CDM or the SIDM-c0.1 model.

With the large sample size we have, higher order differences can be revealed from the distribution of the measured ellipticities. In the left panel of Figure~\ref{fig:eps_cdf}, we show the cumulative distribution function (CDF) of the average ellipticities at $0.1\operatorname{-}0.2\,R_{\rm 200}$. In this domain, the two SIDM models with relatively high cross-sections (SIDM-c1 and SIDM-c0.5) give systematically lower ellipticities than CDM, while the SIDM-c0.1 model is indistinguishable from CDM. In some parts of the CDF, the SIDM-c0.1 model predicts even higher ellipticities than CDM, though we are unable to tell if it is due to a physical effect or purely statistical noise. For the observed samples, the ``non-peaky'' case has a more extended high ellipticity tail than the ``peaky'' case and agrees better with the simulation results in general. Despite the even more extended high ellipticity tail, the SIDM-c1 and SIDM-c0.5 models agree best with the observed ``non-peaky'' sample, while CDM predicts systematically higher ellipticities by about $0.03$ (manifested as the difference in the median values and a global shift in the CDF). However, the difference in the median ellipticity is significantly mitigated by the large sample variations. In the top right panel of Figure~\ref{fig:eps_cdf}, we show the probability distribution function (PDF) of the ellipticities. The PDFs better reveal the features at the tails of the distributions. This comparison also demonstrates that the shift of the CDFs of SIDM-c1 and CDM are not caused by occasional peaks in the PDF driven by statistical noises, but by a real and systemetic global shift in the PDF. Independent of the dark matter model employed, both the low ellipticity ($\lesssim 0.1$) and high ellipticity ($\gtrsim 0.4$) tails of the simulated clusters are missing in the observational samples. However, this could be related to the baryonic physics (e.g. radiative cooling, star formation and stellar/AGN feedback) that are not included in the simulation. An evidence is that the ``non-peaky'' sample in observations (presumably less affected by cooling and star formation) shows much more high-ellipticity clusters than the ``peaky'' sample. The impact of baryonic physics and potential selection biases will be discussed in detail in Section~\ref{sec:discussion}.

\subsection{Non-parametric statistical analysis}
\label{sec:results-likelihood}

{\it (\rmnum{1}) Kolmogorov-Smirnov statistic}: The two-sample Kolmogorov-Smirnov (KS) test is a nonparametric test that compares the (empirical) CDF of two datasets. It measures the likelihood that two univariate datasets are drawn from the same underlying parent probability distribution. Let $x_1,x_2,...,x_{\rm m}$ and $y_1,y_2,...,y_{\rm n}$ be samples of independent observations of populations with continuous distribution functions $F$ and $G$, respectively. The empirical CDFs are $F_{\rm m}$ and $G_{\rm n}$ (i.e. the number of observations $x_{\rm i}$'s which do not exceed $u$ is $m\,F_{\rm m}(u)$ and similarly for $G$). To test the null hypothesis $F=G$, the KS statistic is defined as 
\begin{equation}
    D_{\rm m,n} \equiv \sqrt{\dfrac{m\,n}{m+n}}\, \underset{\rm u}{\rm sup}\,| F_{\rm m}(u) - G_{\rm n}(u) |,
\end{equation}
where $\rm sup$ represents the supremum of the set of distances. The probability distribution $P_{\rm ks}(t) \equiv Pr({D_{\rm m,n}\leq t\,|\,F=G})$ is mathematically proven to be independent of the detailed form of $F$ or $G$, if $F$ and $G$ are continuous. We use the {\sc Scipy} implementation of the two-sample KS test, which follows \citet{Hodges1958} treatment of the probability function $P_{\rm ks}(t)$. The null hypothesis is rejected at the significance level $\alpha$ if $D_{\rm m,n} > K_{\alpha}$, where $K_{\alpha}$ is found from $Pr({D_{\rm m,n}\leq K_{\alpha}\,|\,F=G})=\alpha$. In the following, the value $1-\alpha$ will be referred to as the $p$ value. The $p$ value should be interpreted as {\it the probability of observing an equal or larger discrepancy in the empirical CDFs, $F_{\rm m}$ and $G_{\rm n}$, than what was observed from the data} in the hypothetical context where $F=G$, instead of {\it the probability that the null hypothesis $F=G$ is true}.

For our purpose here, we perform KS tests between the samples of ellipticities measured from simulations and observations. The tests will be performed between simulations of each dark matter model and each observational group, respectively. The null hypothesis is that the simulation and observational samples are randomly drawn from the same underlying distribution of ellipticities. For each test, we obtain the statistical significance $\alpha$ at which this null hypothesis is rejected. In the bottom right panel of Figure~\ref{fig:eps_cdf}, we show the value $p\equiv 1-\alpha$ versus SIDM cross-section. For the ``non-peaky'' sample, the CDM model is rejected at about $90\%$ confidence level, while the SIDM models with $(\sigma/m)\geq 0.5\cpm$ are only constrained at about $40\%$ confidence level, and thus have greater chance of being consistent with the observational sample. On the other hand, for the ``peaky'' sample, even the SIDM-c1 model is rejected at about $90\%$ confidence level. In terms of the KS statistics, CDM appears to be more consistent with the data than the SIDM-c0.1 model. This is due to the fact that the SIDM-c0.1 model predicts even higher ellipticities than CDM in some parts of the CDF, as shown in the left panel of Figure~\ref{fig:eps_cdf}. However, we are unable to tell if this is due to a physical effect or purely statistical noise.

\noindent {\it (\rmnum{2}) Anderson-Darling statistic}: The KS test is most sensitive when the empirical CDFs differ in a global fashion, but could be misleading if there are repeated crossings between the CDFs or the deviations take place at the tails of the distributions. Alternatively, the Anderson-Darling \citep[AD,][]{Anderson1952,Anderson1954} test was designed to overcome these problems and has been proven more sensitive than the KS test with extensive implications. The two-sample AD statistic is defined as~\citep{Darling1957,Pettitt1976,Scholz1987} 
\begin{equation}
    A^{2}_{\rm m,n} = \dfrac{mn}{N}\int_{-\infty}^{\infty}\dfrac{\left[F_{\rm m}(u)-G_{\rm n}(u)\right]^2}{H_{\rm N}(u)\left[1-H_{\rm N}(u)\right]}\,{\rm d}H_{\rm N}(u),
\end{equation}
where $N=m+n$ and $H_{\rm N}(u)= [mF_{\rm m}(u)+nG_{\rm n}(u)]/N$. The weighting term $1/H_{\rm N}(u)\left[1-H_{\rm N}(u)\right]$ gives greater weight to displacements at the tails of the distribution. The probability distribution of the AD statistics has also proven to be independent of the detailed form of $F$ and $G$. For numerical computation and assessment of the statistical significance, we adopt the {\sc Scipy} implementation of the method following \citet{Scholz1987}. Similar to the KS tests, we perform the AD tests between the simulated and the observed samples, and the results are illustrated in the bottom right panel of Figure~\ref{fig:eps_cdf}. Since the table of AD statistics and significance levels in \citet{Scholz1987} only covers $p$ values from $0.1\%$ to $25\%$, the $p$ values we get from this test are capped accordingly, as marked by the arrows in the figure. The AD tests generally give lower $p$ values than the KS tests, which suggest that the models are rejected at a higher confidence level.

\section{Discussions}
\label{sec:discussion}
\subsection{Statistical uncertainties}
\label{sec:boot}

\begin{figure}
    \centering
    \includegraphics[width=0.49\textwidth]{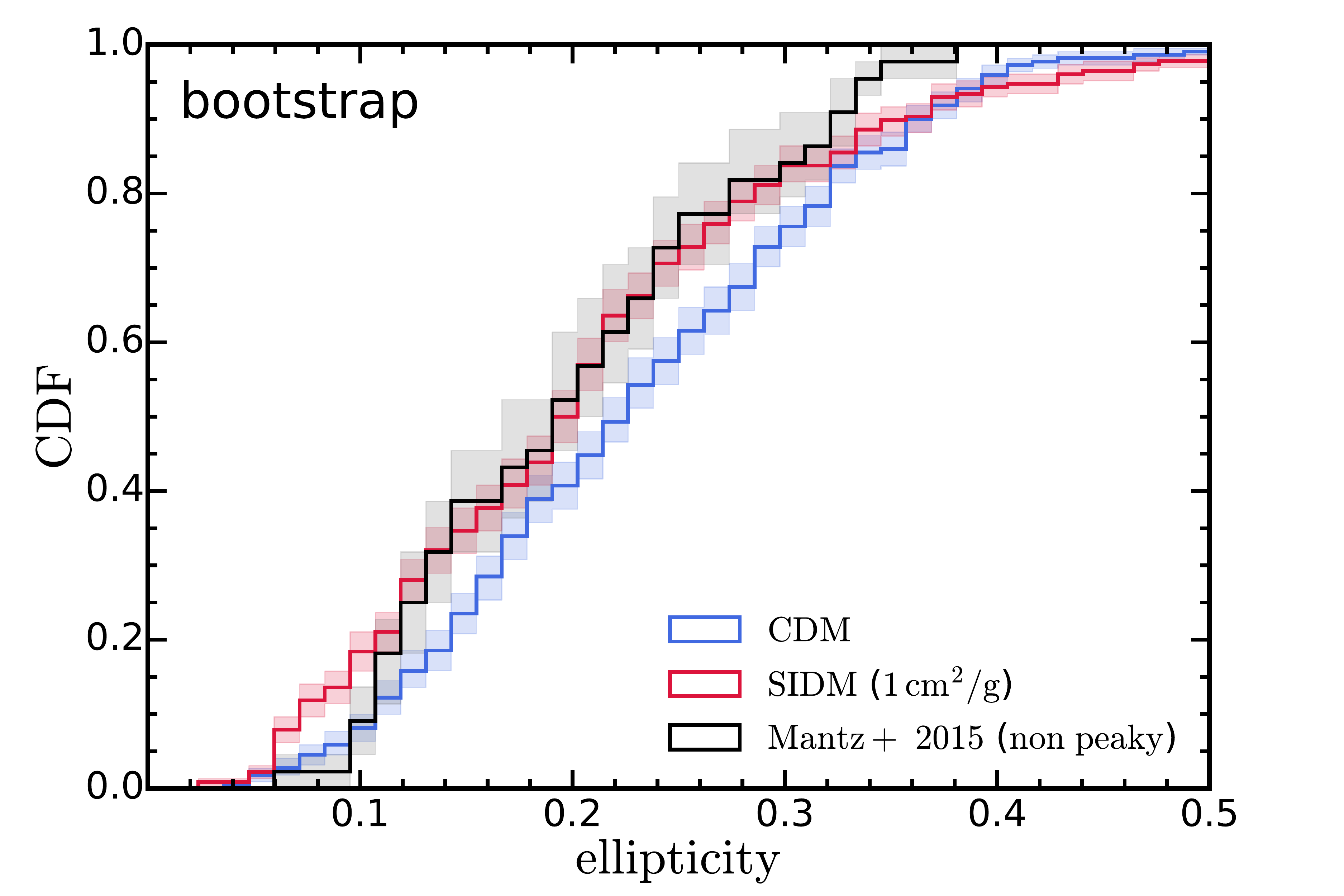}
    \includegraphics[width=0.49\textwidth]{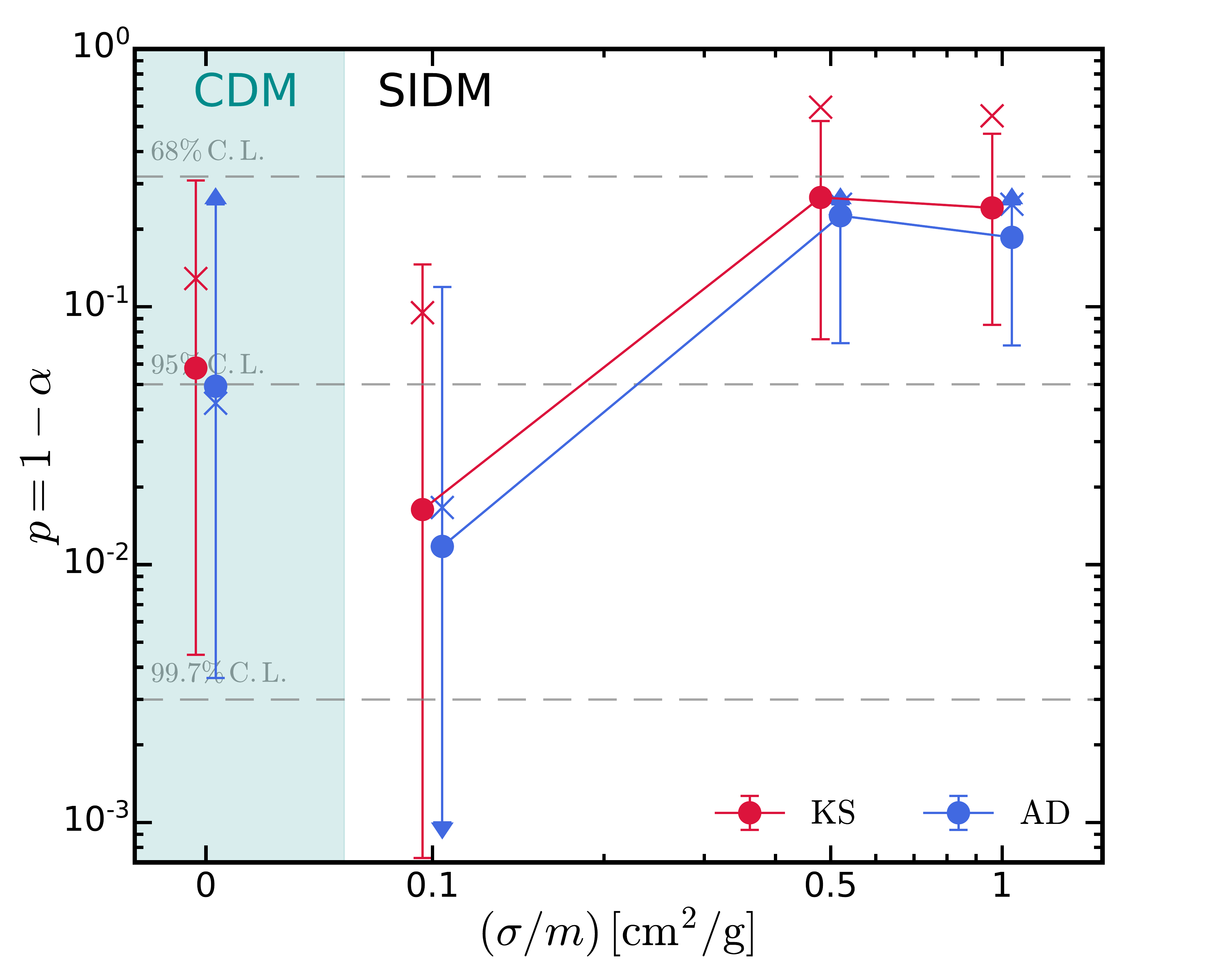}
    \caption{Ellipticity CDFs and statistics from the bootstrapped samples. {\it Top:} Ellipticity CDFs of the bootstrapped samples for the CDM and SIDM-c1 models and the ``non-peaky'' sample from observations. The $1\sigma$ dispersions of the CDFs are shown by the shaded regions. The discrepancy found between CDM and the observed sample is larger than the statistical uncertainties illustrated here. {\it Bottom:} The $p$ values of KS and AD tests for the bootstrapped samples. The median $p$ values and the $1\sigma$ scatters are shown by solid circles with error bars. The $p$ values from the measurements of the original samples are shown by crosses. Since the numerical implementation of the AD test only covers the $p$ values from $0.1\%$ to $25\%$, the bootstrapped results are thus capped, as marked by the arrows in the figure. Even taking into account the scatter in $p$ value, the CDM and SIDM-c0.1 models are rejected with a $68\%$ confidence level.}
    \label{fig:boot}
\end{figure}

To measure the statistical uncertainties from the limited sample size and viewing angle choices, we generate $1000$ bootstrapped realizations for each observational sample and for simulations in each dark matter model. The number is chosen to give a converged assessment of the KS or AD statistics. The bootstrapped realizations of the sample have the same sample size as the original one. In the top panel of Figure~\ref{fig:boot}, we present the median CDFs and the $1\sigma$ dispersions from the bootstrapped samples of the simulated CDM and SIDM-c1 samples and the observed ``non-peaky'' sample. The displacement between CDM and the observed sample is robust against the statistical uncertainties measured here. The SIDM-c1 model and its bootstrapped realizations are in good agreement with the observed sample, particularly around the center of the distribution, though with more extended tails at both ends of the distribution. In the bottom panel of Figure~\ref{fig:boot}, the median $p$ values and $1\sigma$ dispersions of the KS and AD statistics are shown. The statistics have been computed for each pair of the bootstrapped realizations, leading to in total one million measurements. For the KS test, the $p$ values of the bootstrapped samples are systematically lower than that of the original sample, and in some cases the original value even lies outside the $1\sigma$ scatter. This would be expected when the CDFs of the original sample differ in a global fashion\footnote{A small displacement from this state in bootstrapping will more likely lead to a larger KS statistic $\underset{\rm u}{\rm sup}\,|F-G|$ (thus lower $p$ value) and a shift of the location where the maximum is reached. Stated in another way, a lower $p$ value (a higher KS statistic) corresponds to a larger number of realizations of bootstrapped samples (and thus a larger entropy). So this can be understood as the entropy gain when deviating from a (quasi-)equilibrium state.} ($|F-G|$ weakly depends on ellipticity). On the other hand, the median $p$ value of the AD tests of bootstrapped samples agrees well with the original measurement. Neglecting the cap at $p=0.1\%$ and $p=25\%$, the bootstrapped results of the KS and AD tests agree remarkably well and both of them suggest that the CDM and SIDM-c0.1 models are rejected at $68\%$ confidence level, as conservative estimates.

\subsection{Selection bias and systematic uncertainties}

\begin{figure}
    \centering
    \includegraphics[width=0.49\textwidth]{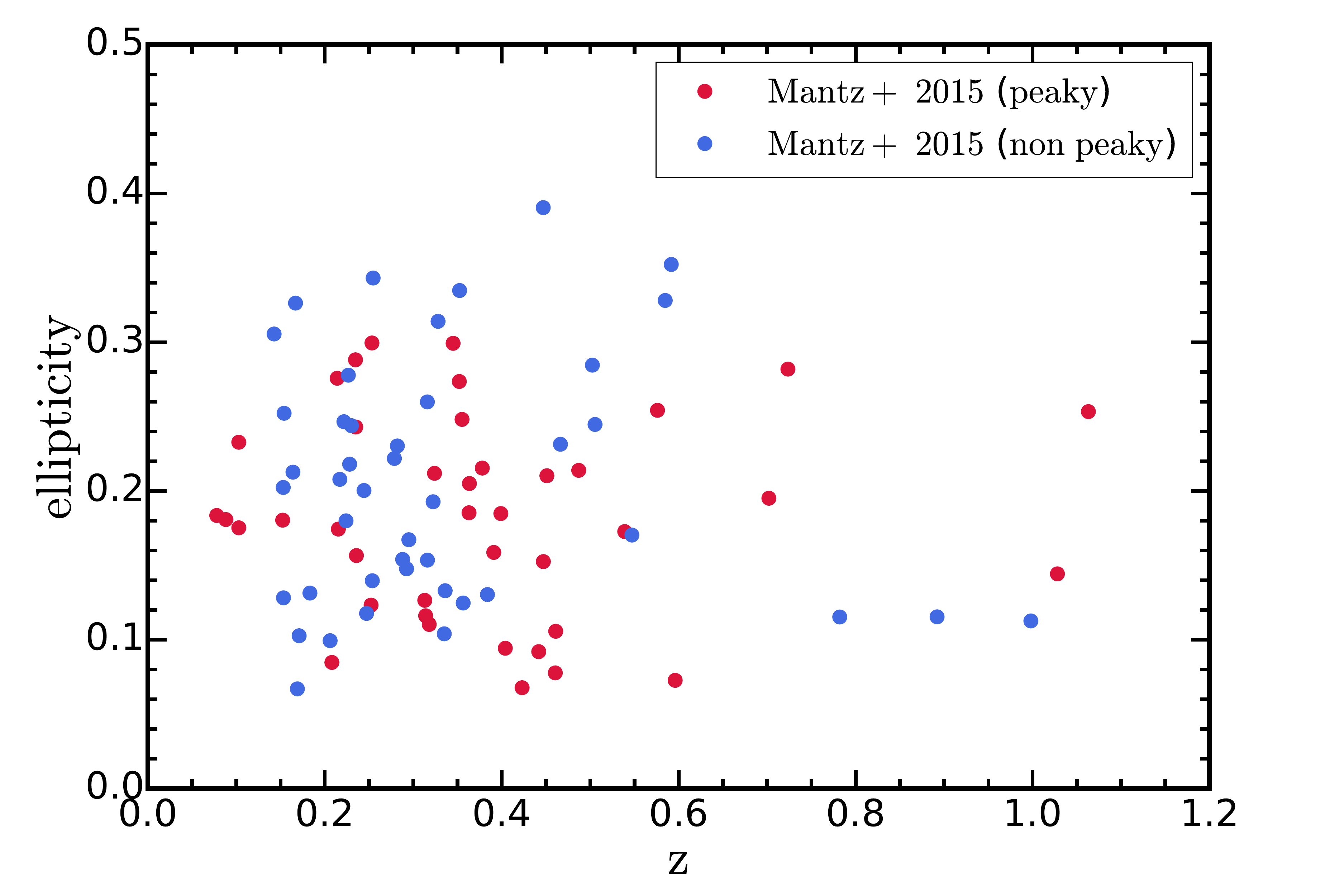}
    \includegraphics[width=0.49\textwidth]{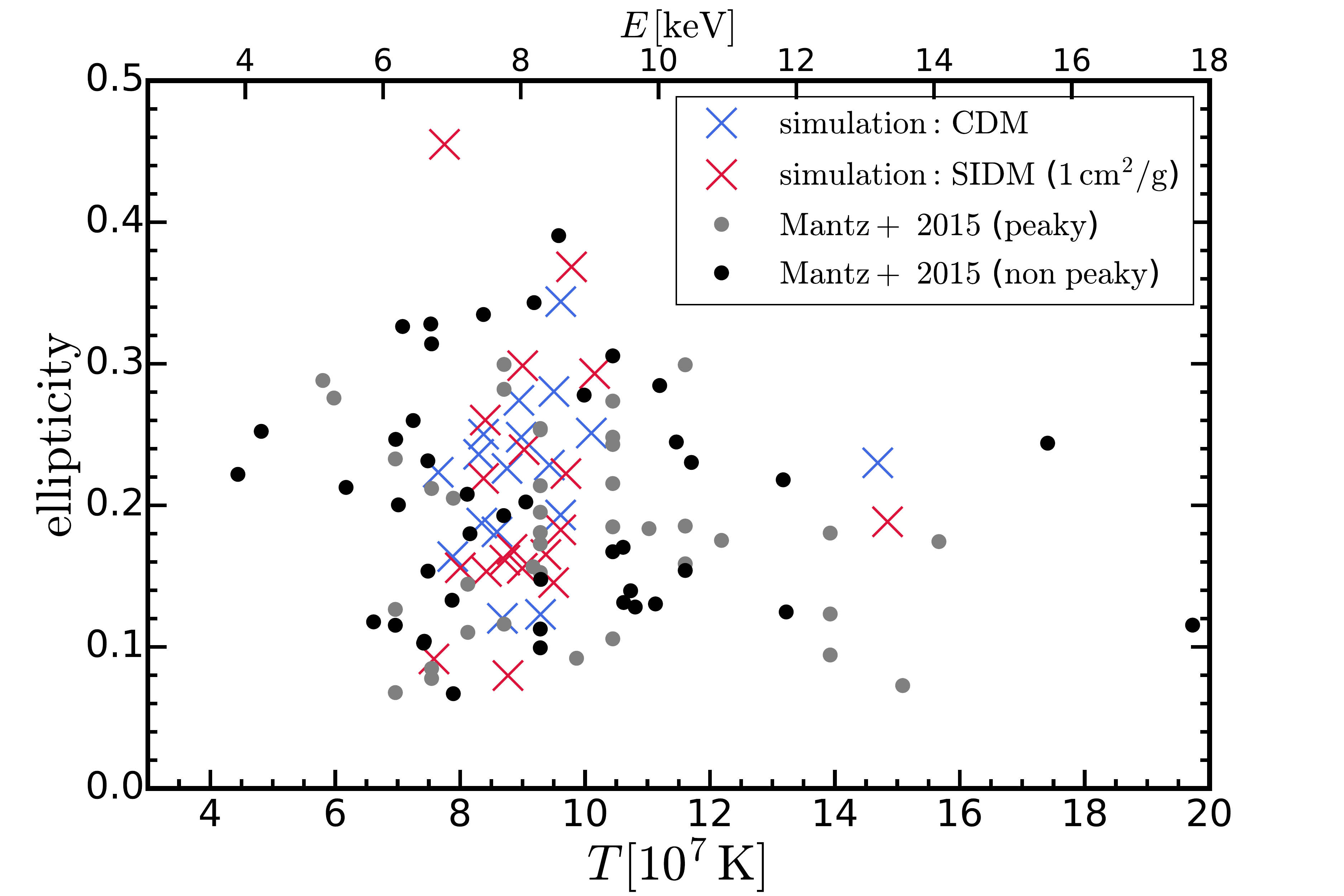}
    \caption{{\it Top:} Ellipticity versus redshift of the observed clusters. The ellipticity does not show any apparent dependence on redshift. Note that these clusters have been pre-selected as dynamically relaxed objects through the SPA criteria. {\it Bottom:} Ellipticity versus temperature. No apparent dependence on temperature is found either. However, the simulated clusters have a narrower distribution in temperature.}
    \label{fig:e_vs_z_and_T}
\end{figure}

\begin{figure}
    \centering
    \includegraphics[width=0.49\textwidth]{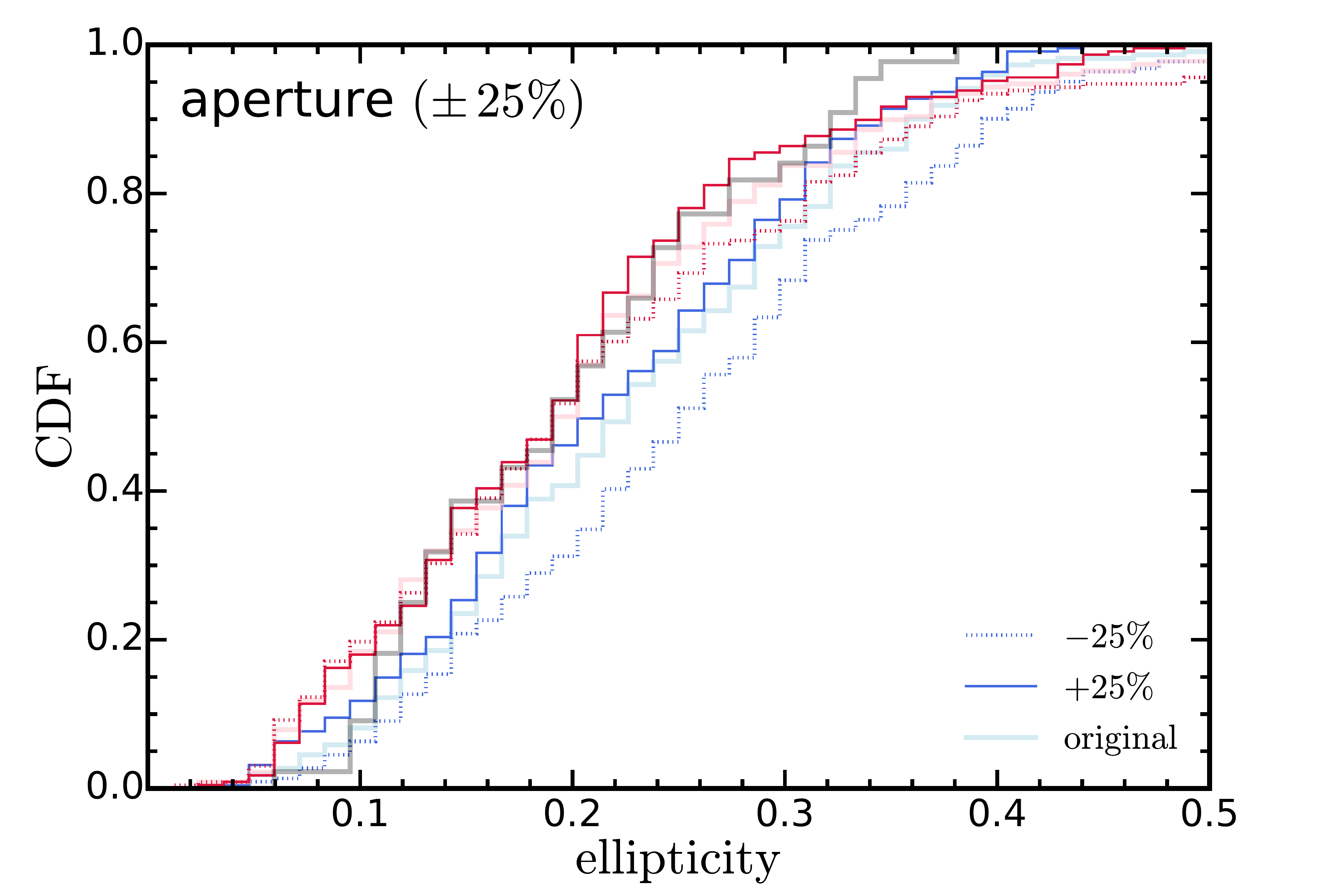}
    \includegraphics[width=0.49\textwidth]{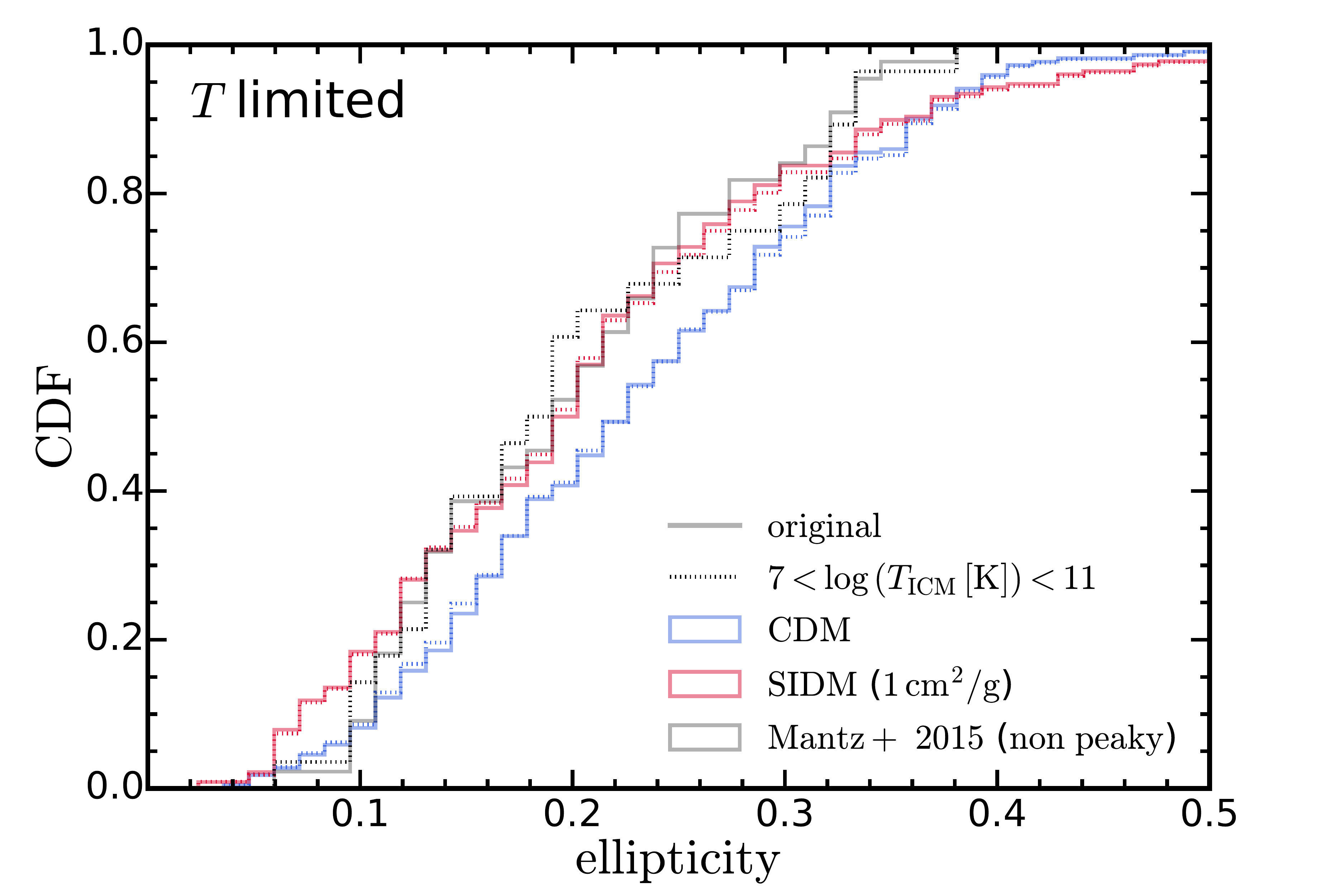}
    \includegraphics[width=0.49\textwidth]{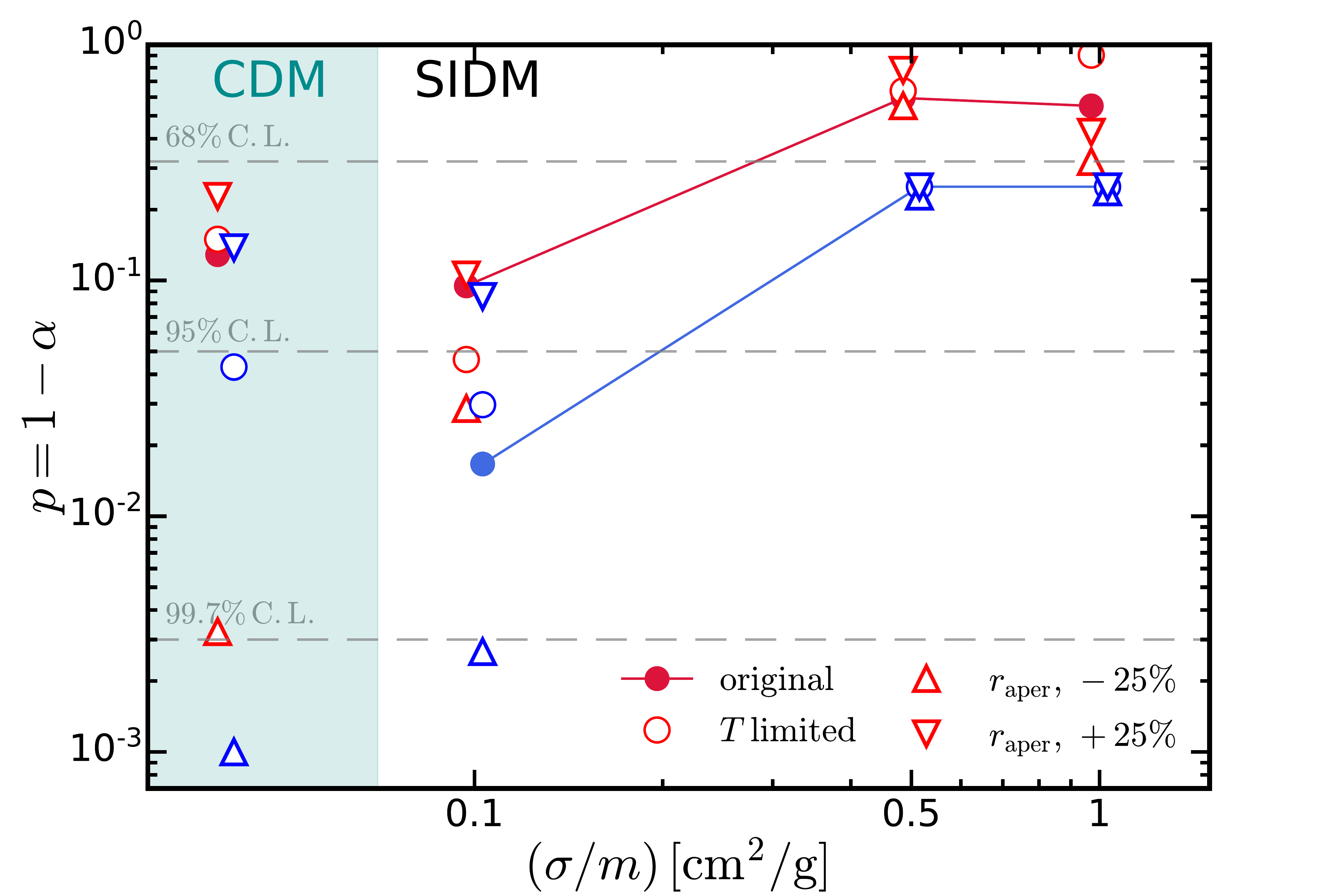}
    \caption{{\it Top:} Ellipticity CDFs when varying the radial aperture of the measurements. The CDF when increasing (decreasing) the radial aperture by $25\%$ is shown as the solid (dotted) lines. The original CDMs are shown as the transparent lines. The comparison demonstrates that the results are robust against aperture shifts. {\it Middle:} Ellipticity CDFs when excluding outliers in the ICM temperature distribution. The CDFs of the temperature limited samples are presented in dashed lines while those of the original samples are shown in solid lines. The impact of the temperature outliers is small. {\it Bottom:} The $p$ value of KS and AD tests when varying the radial aperture of the measurements or applying a temperature selection criterion. The conclusion that the CDM and dSIDM-c0.1 models are disfavored at $68\%$ confidence level is not altered by either the aperture shift or the temperature selection criterion.}
    \label{fig:testbias}
\end{figure}

Our analysis is potentially subject to selection biases in cluster redshift and temperature. In the top (bottom) panel of Figure~\ref{fig:e_vs_z_and_T}, we show the ellipticity versus cluster redshift (temperature). The ICM temperatures of the simulated clusters are approximated as the virial temperature. Although no obvious redshift or temperature dependence is found for the measured ellipticity, the simulated clusters have a narrower temperature distribution. This is likely related to the selection criteria for these clusters from the parent large-box simulation. They are the most massive haloes that are classified as dynamically relaxed and have not undergone recent mergers. Clusters of even higher masses are most likely perturbed by recent mergers.

As illustrated in Figure~\ref{fig:eps_profile}, the ellipticity of the isophotes of the simulated clusters exhibit a weak radial dependence, an effect that is more apparent at small radii and in CDM. The observed samples show a similar trend as well, albeit with the ``non-peaky'' clusters displaying a much stronger radial dependence. In the top panel of Figure~\ref{fig:testbias}, we show the CDF of ellipticities measured at slightly smaller/larger ($\pm 25\%$) radii as solid/dotted lines. The ellipticities measured at smaller radii are typically larger. As expected, the ellipticity CDF in CDM is more affected by the aperture than in the SIDM-c1 model, but their systematic difference is robust against the shift of the aperture. In the middle panel of Figure~\ref{fig:boot}, we test the results against the scatter in cluster temperature (or equivalently cluster mass). As shown in the right panel of Figure~\ref{fig:ztdis}, the observational samples have larger scatters in the temperature distribution than the simulated ones and a few hot cluster outliers. To test if these outliers would affect the ellipticity measurements, we limit the analysis to clusters with $7\times 10^{7} \leq T_{\rm vir}\leq 11\times 10^{7}\,{\rm K}$ and show the results as dashed lines in the figure. The results are robust against these outliers. In the bottom panel of Figure~\ref{fig:testbias}, we show the KS and AD statistics after applying the aperture shift or the temperature cut. None of the conclusions we drew in the previous sections is affected by these variations.

Another potential bias originates from the different definitions (and selection criteria) for ``relaxed'' clusters in simulations and observations. The sample of relaxed haloes for zoom-in simulations were selected based on the virial ratio, center offset and subhalo mass fraction~\citep{Brinckmann2018}, which are expected to inherit some intrinsic bias from the morphologically selected observed samples. In the future, constructing a volume-limit sample of massive haloes from large-volume, hydrodynamical simulations would be an important future follow-up project. This would allow a morphology selection processes based on mock X-ray images from simulated clusters, which is fully consistent with the observed sample. However, this certainly requires significantly higher computational cost and more development in the sub-grid models for cluster physics. 

\subsection{Impact of baryonic physics}
\label{sec:diss_baryon}

The response of cluster morphology to baryonic physics is not yet fully understood. Radiative cooling of the intracluster plasma results in a condensed, rotating gas disk in the central part of the halo, fueling star formation and subsequent AGN activity in the Brightest Cluster Galaxy (BCG). The rotation support (and other non-thermal processes in the intracluster plasma) breaks hydrostatic equilibrium and the flattening of the three-dimensional gas distribution is reflected by the larger ellipticities of two-dimensional isophotes. \citet{Fang2009} found that the ellipticities of X-ray isophotes are enhanced at small cluster-centric radii, $r\lesssim 0.4\,R_{\rm 500}\sim 0.26\,R_{\rm 200}$\footnote{Assuming an NFW profile with concentration $c=4$, a typical value for cluster-mass haloes, $R_{\rm 500} \simeq 0.65 R_{\rm 200}$ and $R_{\rm vir}(z=0) \simeq R_{\rm 100} \simeq 1.35 R_{\rm 200}$, where the second argument assumes the redshift-dependence from \citet{Bryan1998} for $\Delta_{\rm c}(z=0)\simeq 100$.}, in simulations with radiative cooling and star formation (CSF) compared to adiabatic runs. The average ellipticity over sightlines reached $0.6$ at $r\sim 0.1\,R_{\rm 500}$ ($\sim 0.065\,R_{\rm 200}$) as opposed to $0.3$ in adiabatic runs. A similar phenomenon was found by \citet{Lau2011} between non-radiative (NR, i.e. adiabiatic) cooling runs and CSF runs. However, the flattening of the isophotes due to cooling was confined to smaller radii $r\lesssim 0.1\,R_{\rm 500}$. 

On the other hand, at the radii beyond the scale of the central gas disk, haloes in CSF runs were more spherical than those in adiabatic models. For example, isophote ellipticities were lower at $r\gtrsim 0.1\operatorname{-}0.2\,R_{\rm 500}$ ($0.065\operatorname{-}0.13\,R_{\rm 200}$) in CSF simulations with respect to those in NR simulations by about $0.1$ (see \citealt{Lau2011} and also \citealt{Fang2009}, noting that the latter found a similar difference, but at larger radii). Similar effects were also found in \citet{Battaglia2012} and \citet{Suto2017} out to half of the virial radius with about $0.05$ difference in two-dimensional axis-ratios. The azimuthal scatter of surface brightness was found to be substantially lower in cool-core clusters in observations and CSF simulations compared to adiabatic runs \citep{Eckert2012}, suggesting rounder distributions of gas. The shapes of the gas (and dark matter) distributions are sensitive to the degree of the central concentration of the total mass. As intracluster gas cools and flows towards the halo center, the distribution becomes more spherical \citep[e.g.,][]{Dubinski1994, Evrard1994, Tissera1998, Kazantzidis2004,Debattista2008, Suto2017, Shen2021}. In this study, the ``peaky'' clusters have lower isophote ellipticities than the ``non-peaky'' clusters, which is consistent with this picture. The fact that the ``non-peaky'' clusters in observations still have more concentrated surface brightness profiles than the simulated ones indicate some level of cooling even in the ``non-peaky'' sample that is not captured by the adiabatic simulations. This effect has the potential to make the CDM results presented in this paper more consistent with observations.

Meanwhile, the cooling and condensation of gas can feed both star formation and accretion onto supermassive black holes (SMBHs) harboured by the BCG. The resulting stellar/supernovae and AGN feedback can inject substantial amounts of energy into the ICM through radiation, kinetic outflows and power jets of relativistic particles. As important heating mechanisms, they can compensate the energy loss due to radiative cooling and mitigate the sphericalizing effect of cooling. In addition, potential anisotropic feedback processes (e.g., bi-modal jets, bubbles, outflows from satellite galaxies) can disturb the ICM and create non-thermal pressure support for the intracluster gas in certain directions, further breaking the sphericity of the halo. In numerical simulations, \citet{Battaglia2012} and \citet{Suto2017} found that clusters are less spherical when AGN feedback is included relative to including only radiative cooling. As an enlightening attempt, \citet{Robertson2018} performed a series of galaxy cluster simulations that includes baryonic physics and found diverse density profiles of cluster-mass haloes, which can be understood in terms of their different final baryon distributions. This was followed by BAHAMAS--SIDM simulations \citep{Robertson2019}, which is the first large-volume cosmological set of simulations including both SIDM and baryonic physics, including AGN feedback. Although considerable differences were found in the shape of dark matter distributions, the discrepancy is weakened by baryonic effects and were not reflected in the distribution of gas or stars within galaxy clusters. However, there is no consensus yet on the strength and underlying mechanism of AGN feedback as well as its numerical implementation. And the numerical challenge to resolve baryonic physics processes for the large simulation volumes required to sample massive clusters still exist. It is still hard to tell whether the baryonic physics that primarily influence the central part of the clusters, and which are not present in our simulations, can explain the discrepancy we report here between adiabatic CDM simulations and observations of clusters.

\section{Conclusions}
\label{sec:conclusion}

In this paper, we study the X-ray morphology of massive, dynamically relaxed clusters based on a suite of cosmological hydrodynamical zoom-in simulations of $19$ haloes with $M_{200}\simeq 1\operatorname{-}2 \times 10^{15}\msun$, simulated in CDM and SIDM models with three different (constant) cross sections per unit mass: $(\sigma/m)= 0.1, 0.5$ and $1.0\cpm$. The structural properties of both dark matter and intracluster gas in these clusters are studied quantitatively in detail. These simulations include adiabatic gas of which the X-ray emission is modelled to create mock soft X-ray images. We perform ellipse fitting on the isophotes at intermediate radii of the clusters and compare the ellipticities with those measured from real cluster X-ray images. Our findings can be summarized as follows.

\begin{itemize}
    \item The intracluster gas in the adiabatic simulations is in almost perfect hydrostatic equilibrium until reaching the hydro resolution limit. The gas temperature within $0.1\,R_{\rm 200}$ is slightly lower in SIDM with increasing cross-sections. Although the central dark matter density profile in SIDM is distinct from that in CDM (when $(\sigma/m)\geq 0.1\cpm$ as tested by our simulation suite), the gas density profiles of the two cases are almost indistinguishable down to the resolution limit. 
    
    \item Similar to what was found in \citet{Brinckmann2018}, the three-dimensional shapes of the dark matter distribution in CDM and SIDM-c1 exhibit at least $2\sigma$ level discrepancy out to large cluster-centric radii ($r\sim 0.2\,R_{\rm 200}$). For all the models, the gas distributions are systematically more spherical than those of dark matter, as a consequence of gas in hydrostatic equilibrium tracing the isopotential surfaces, which are more spherical than the mass distribution. The variation in axial ratios decreases to about $1\sigma$ level at $r\sim 0.1\operatorname{-}0.2\,R_{\rm 200}$ between CDM and SIDM-c1.
    
    \item The surface brightness profiles in SIDM are remarkably similar to those produced in CDM. Both of them are in good agreement with observations at the outskirts of the clusters ($r\gtrsim 0.1\,R_{\rm 200}$), while the observed clusters develop cuspy profiles at the center, especially for the selected cool-core (``peaky'') clusters. 
    
    \item Two-dimensional shape analysis is performed on the real and mock X-ray images, of which the isophotes at the target radius ($r\gtrsim 0.1\operatorname{-}0.2\,R_{\rm 200}$) are fitted with ellipses. We find that the ellipticities of the observed ``non-peaky'' clusters are systematically lower than the CDM prediction, and interestingly in good agreement with the SIDM models with $(\sigma/m)\geq 0.5\cpm$. Based on statistical tests of the bootstrapped samples, we find that the CDM and SIDM-c0.1 models are conservatively disfavored at $68\%$ confidence level. The result is robust against aperture choices and selection biases in cluster temperatures and redshifts.
    
\end{itemize}

In conclusion, we demonstrate that the X-ray morphology of massive, relaxed clusters is a promising channel to constrain dark matter self-interactions. Even though the dark matter model-dependent variations in shape is smaller in the gas distribution and weakened by projection effects, distinct signals can be identified with a large sample of observed and simulated clusters. Our analysis favors SIDM models with relatively high cross-sections. However, effects due to baryonic physics, including cooling, star formation and feedback effects that are not captured by our adiabatic simulations is the primary source of uncertainty, and has the potential to reconcile simulations with observations within the CDM framework. Follow-up cluster simulations with full baryonic physics are required to confirm our findings.

\section*{Acknowledgements}
The computations in this paper were run on the Faculty of Arts and Sciences Research Computing (FASRC) Cannon cluster supported by the FAS Division of Science Research Computing Group at Harvard University. This research made use of {\sc Photutils}, an {\sc Astropy} package for detection and photometry of astronomical sources \citep{larry_bradley_2020}. TB was supported by the United States Department of Energy (DOE) grant DE-SC0017848, through the Istituto Nazionale di Fisica Nucleare of Italy (INFN) project GRANT73/Tec-Nu, and from the COSMOS network (www.cosmosnet.it) through the Italian Space Agency (ASI) Grants 2016-24-H.0 and 2016-24-H.1-2018. DR acknoledges support by the National Aeronautics and Space Administration (NASA) under award number NNA16BD14C for NASA Academic Mission Services. MV acknowledges support through NASA Astrophysics Theory Program (ATP) 19-ATP19-0019, 19-ATP19-0020, 19-ATP19-0167, and the National Science Foundation (NSF) grants AST-1814053, AST-1814259, AST-1909831, AST-2007355 and AST-2107724. JZ acknowledges support by a Grant of Excellence from the Icelandic Research fund (grant number 206930). SA acknowledges support from the United States Department of Energy under contract number DE-AC02-76SF00515.

\section*{Data Availability}
The simulations in this paper were run on the super-computing system Cannon at Harvard University and the data were stored on the Engaging cluster at Massachusetts Institute of Technology. The data underlying this article can be shared on reasonable request to the corresponding author.









\bsp	
\label{lastpage}
\end{document}